\documentclass[prd,superscriptaddress,nofootinbib,preprintnumbers]{revtex4}
\usepackage{amsmath,amsfonts,amssymb}
\usepackage{graphicx}
\usepackage{dcolumn}
\usepackage[mathlines]{lineno}
\usepackage{float}
\usepackage{hyperref}
\usepackage[usenames,dvipsnames]{color}
\usepackage{etex}
\usepackage{booktabs}
\usepackage{adjustbox}
\usepackage{xspace}
\usepackage{rotating}
\usepackage{longtable}
\usepackage{multirow}
\usepackage{booktabs}
\usepackage{array,tabularx,epsfig,mathrsfs}
\usepackage{ifthen}
\usepackage{amsfonts}
\usepackage{bbold}
\usepackage{mathtools}

\begin{document}

\title{Radiative neutrino mass generation and dark matter through vectorlike leptons}

\author{Mohamed Amin Loualidi}
\email{ma.loualidi@uaeu.ac.ae}
\author{Salah Nasri}
\email{snasri@uaeu.ac.ae, salah.nasri@cern.ch}
\affiliation{Department of Physics, United Arab Emirates University, Al-Ain, UAE}

\author{Maximiliano A. Rivera}
\email{maximiliano.rivera@usm.cl}
\affiliation{Departamento de F\'isica, Universidad T\'ecnica Federico Santa Mar\'ia, Casilla 110-V, 
 Valparaiso, Chile}

\begin{abstract}This study presents a radiative three-loop model for neutrino mass generation, employing an asymmetric Yukawa coupling between two new scalar $SU(2)_L$ doublets and vectorlike lepton doublets. Dark matter candidates arise from one of the scalar doublets and contribute to neutrino mass generation through the mass splitting between its neutral components. The singly charged scalars are also essential for neutrino mass, with the charged states of the two doublets mixing with one another. A single generation of vectorlike leptons yields two nonzero neutrino masses as a consequence of the asymmetric Yukawa combinations entering the neutrino mass matrix. The model is tested against dark matter phenomenology, neutrino mass and mixing data, and the charged lepton flavor-violating process $\mu \rightarrow e \gamma$, showing compatibility with current bounds and leading to experimentally accessible predictions.
\end{abstract}
%%%%%%%%%%%%%%%%%%%%%%%%%%%%%%%%%%%%%%%%%%%%%%%%%%%%%%%%%%%%%%%%%%
%%%%%%%%%%
%\pacs{ {pacs needed}}}

\maketitle

%%%%%%%%%%%%%%%%%%%%%%%%%%%%%%%%%%%%%%%%%%%%%%%%%%%%%%%%%%%%%%%
\section{Introduction} 
\label{sec1}
%%%%%%%%%%%%%%%%%%%%%%%%%%%%%%%%%%%%%%%%%%%%%%%%%%%%%%%%%%%%%%
Neutrino oscillation experiments have established that neutrinos possess nonzero masses \cite{Super-Kamiokande:1998kpq,SNO:2002tuh}, pointing to physics beyond the Standard Model (SM). Astrophysical and cosmological observations likewise provide strong evidence for dark matter (DM), which the SM cannot account for \cite{Bertone:2004pz}. Elucidating the mechanism underlying neutrino mass generation and identifying the particle nature of DM thus constitute two of the most pressing open questions in contemporary particle physics (see Ref. \cite{Mohapatra:2005wg} for a review). Radiative neutrino mass generation offers an elegant approach that generates neutrino masses through loop-level quantum corrections, naturally suppressing their scale mediated by the DM candidate. A paradigmatic example is the one-loop model proposed by Ma \cite{Ma:2006km}, linking neutrino physics and DM phenomena through a $Z_2$ symmetry. Higher-loop realizations, that break the lepton number by two units $\Delta L= 2$, are studied in \cite{Babu:2001ex, deGouvea:2007qla}, focusing on higher-dimensional operators involving only SM quarks, leptons, and the Higgs doublet \cite{Cai:2017jrq, Krauss:2002px, Aoki:2008av, Cheung:2017efc, Kajiyama:2013lja, Ahriche:2014cda, Ahriche:2014oda, Hatanaka:2014tba, Nishiwaki:2015iqa, Okada:2015hia, Ahriche:2015wha, Cepedello:2020lul}.

Despite existing classifications, there remains a gap in the comprehensive coverage of all $\Delta L = 2$ operators, particularly those that incorporate SM gauge fields, as extensively discussed in the literature \cite{Chen:2006vn,delAguila:2011gr, Gustafsson:2012vj,Alcaide:2017xoe, Jin:2014glp}. An analytical overview from the perspective of effective field theory regarding these operators is provided in Refs. \cite{Gustafsson:2014vpa,Gustafsson:2020bou}, which include several illustrative examples of the respective models and their associated phenomenology. In the study conducted by Chen et al. \cite{Chen:2014ska}, a novel model is introduced in the context of the three-loop extension of the Krauss, Nasri, and Trodden model \cite{Krauss:2002px}. This model is not included in the classification presented by \cite{Gustafsson:2020bou}. Nevertheless, the effective operator utilized by this new model remains consistent with those that integrate SM gauge fields. \\

In this study, we develop a theoretical framework building on the foundational work of Chen et al. \cite{Chen:2014ska}. Our model incorporates vectorlike leptons (VLLs) through an asymmetric Yukawa interaction that couples the VLLs to SM leptons as well as to newly introduced $SU(2)_L$ scalar doublets carrying distinct hypercharges. Neutrino mass generation is facilitated at the three-loop level through DM interactions, which are mediated by nonzero scalar interaction between {\it CP}-odd and {\it CP}-even fields, alongside a mixing involving singly charged fields. A distinctive feature of our construction is that a single generation of VLLs suffices to generate a rank-2 neutrino mass matrix, $m^\nu_{ab} \propto (g_a \tilde{g}_b + \tilde{g}_a g_b)$. Consequently, two neutrino masses are nonzero, while the third remains massless, providing a minimal and predictive realization of the observed neutrino mass spectrum without enlarging the fermion sector. We explore the viability of the model exploring the DM sector, neutrino mass and mixing angles, and charged lepton flavor violation processes, demonstrating successful accommodation of these constraints while offering testable predictions. \\

The rest of this paper is organized as follows. Section~\ref{sec:model} details the model and the three-loop neutrino mass matrix as well as the scalar sector of the model. In Sec.~\ref{sec:pheno}, we discuss the phenomenological implications of the model. Our numerical analysis and results are presented in Sec.~\ref{sec:numres}, and conclusions are given in Sec.~\ref{sec:Conclusions}.
%%%%%%%%%%%%%%%%%%%%%%%%%%%%%%%%%%%%%%%%%%%%%%%%%%%%%%%%%%%
\section{The Model}
\label{sec:model}
%%%%%%%%%%%%%%%%%%%%%%%%%%%%%%%%%%%%%%%%%%%%%%%%%%%%%%%%%%%
%
The model presented here was first introduced in \cite{Chen:2014ska} as an ultraviolet completion to the earlier work by Krauss, Nasri, and Trodden (KNT)~\cite{Krauss:2002px}. It explores an extension of the SM that includes two scalar fields, both transforming as $SU(2)_L$ doublets but carrying different hypercharges—specifically $S_1\sim (1,2,1)$ and $S_2\sim (1,2,3)$. The model also introduces a VLL doublet expressed as $\mathcal{F}=\mathcal{F}_L+\mathcal{F}_R\sim (1,2,-1)$. All these newly introduced fields are characterized by odd parity under a $Z_2$ symmetry which ensures that the scalar doublets are inert. As a result, the Lagrangian describing the VLL interactions is expressed as follows:
\begin{eqnarray}
    \mathcal{L} &\supset& -\frac{1}{2} M_{\mathcal{F}_i}~\overline{\mathcal{F}_{L_i}} \mathcal{F}_{R_i} -g_{ia}\overline{\mathcal{F}_{L_i}}\, S_1~\ell_{R_a} -\tilde{g}_{ib}~\overline{\ell_{R_b}^{c}}~\tilde{S_2}^\dagger\,\mathcal{F}_{R_i}, 
    \label{eq:lagrangian_yuk}
\end{eqnarray}
where $\tilde{S_2} = i \sigma_2 S_2^{*}$ and $\overline{\mathcal{F}_L} M_\mathcal{F} \mathcal{F}_R$ is the mass term for the VLL. $g_{ia}$ and $\tilde{g}_{ib}$ are the Yukawas of new physics. Since a general model allows for $n$ generations of VLL doublets, here we work with one generation ($n=1$) of VLL for simplicity. Given the particle content of the model, the Lagrangian $\mathcal{L}$ includes terms leading to lepton number violation (LNV) by two units ($\Delta L = 2$) which is essential for neutrino mass generation\footnote{For example, assuming $L(S_1) = 0$ and $L(S_2) = 0$, LNV comes from the $\tilde{g}_{ib}$ term.}.

The particle content of this model given by the SM Higgs doublet $\Phi$, $S_1$, and $S_2$, as well as the VLL doublets, is written explicitly as
\begin{eqnarray}
\Phi =
\left(
\begin{array}{c}
G^+ \\
\frac{1}{\sqrt{2}}(\upsilon + h + i G^0 )
\end{array}
\right),
\qquad
S_1 =
\left(
\begin{array}{c}
S_1^+ \\
\frac{1}{\sqrt{2}}(H + i A )
\end{array}
\right),
\qquad
S_2 =
\left(
\begin{array}{c}
S_2^{++} \\
S_2^+
\end{array}
\right),
\qquad
\mathcal{F} =
\left(
\begin{array}{c}
N_{L,R} \\
E_{L,R}
\end{array}
\right),
\label{eq:fields_convention}
\end{eqnarray}
where the left- and right-handed components of the VLL doublet are written explicitly, in accordance with the chiral structure of the interactions in the Lagrangian of Eq.~(\ref{eq:lagrangian_yuk}). This Lagrangian may then be expanded in terms of the component fields defined in Eq.~(\ref{eq:fields_convention}) as
\begin{eqnarray}
\mathcal{L} &\supset& -\frac{1}{2} m_N \overline{N_{L_i}}N_{R_i} 
-\frac{1}{2} m_E \overline{E_{L_i}}E_{R_i} -g_{ia}\,\overline{N_{L_i}}\,\ell_{R_a}\,S_1^+
   -g_{ia}\,\overline{E_{L_i}}\,\ell_{R_a}\,(H+iA_0)/\sqrt{2} \nonumber \\
 & & -\tilde{g}_{ib}\,\overline{\ell_{R_b}^{c}}\, N_{R_i} S_2^+
 +\tilde{g}_{ib}\,\overline{\ell_{R_b}^{c}}\, E_{R_i}\,S_2^{++}
\label{eq:mass_yuk_1}
\end{eqnarray}
Throughout this work we consider the minimal realization of the model with a single generation of VLLs. In this setup the neutrino mass matrix is necessarily of rank two, leading to one massless light neutrino. Extensions with additional VLL generations or extra lepton-number-violating contributions would generically lift this feature and generate a nonzero lightest neutrino mass, at the price of reduced predictivity.
%%%%%%%%%%%%%%%%%%%%%%%%%%%%%%%%%%%%%%%%%%%%%%%%%%%%%%%%%%%%%%%%%%%
\subsection{Scalar potential and mass spectrum}
\label{sec:masses}
%%%%%%%%%%%%%%%%%%%%%%%%%%%%%%%%%%%%%%%%%%%%%%%%%%%%%%%%%%%%%%%%%%%
The scalar sector of the model, defined in Eq.~(\ref{eq:fields_convention}), includes the SM Higgs doublet $\Phi$ and new doublets $S_1$ and $S_2$. The general renormalizable gauge and $Z_2$-invariant Higgs potential is given by
\begin{eqnarray}
\mathcal{V}(\Phi,S_1,S_2) &=& -\mu_\Phi^2 |\Phi|^2 + \lambda_\Phi |\Phi|^4 + \mu^2_1 |S_1|^2 + \mu^2_2|S_2|^2 + \lambda_1 |S_1|^4 + \lambda_2 |S_2|^4 + \lambda_3 |S_1|^2 |\Phi|^2 \nonumber \\ &+& \lambda_3^{\prime} (\Phi^\dagger S_1)(S_1^\dagger \Phi) + \lambda_4 |S_2|^2 |\Phi|^2 + \lambda_4^{\prime} (\Phi^\dagger S_2)(S_2^\dagger \Phi) + \lambda_S|S_1|^2 |S_2|^2  \\ &+& \lambda_S^{\prime} (S_1^\dagger S_2)(S_2^\dagger S_1)  
+ (\frac{\lambda_5}{2}(S^\dagger_1 \Phi)^2 + \lambda_7 S^\dagger_2 \Phi\tilde{\Phi}^\dagger S_1+\mathrm{h.c.}), \nonumber
\label{eq:higgs_potential}    
\end{eqnarray}
where $\tilde{\Phi} = i \sigma_2 \Phi^{*}$. To ensure perturbativity, the couplings in the scalar potential and the Yukawa couplings must satisfy the following conditions
\begin{equation}
  |\lambda_i| \leq 4\pi , \quad |g_{ia}|, |\tilde{g}_{ib}| \leq \sqrt{4\pi}.
  \label{perturbativity}
\end{equation}
Here, $\lambda_i$ represents the quartic couplings, while $g_{ia}$ and $\tilde{g}_{ib}$ are the new Yukawa couplings as defined in Eq.~(\ref{eq:lagrangian_yuk}). Additionally, to guarantee that the potential remains stable and bounded from below, it is essential to impose vacuum stability conditions. Applying the copositivity criterion \cite{Kannike:2012pe}, we derive the following constraints on the quartic couplings:
\begin{eqnarray}
    \lambda_\Phi > 0, \quad  \lambda_1 > 0, \quad  \lambda_2  > 0  , \lambda_3 + \lambda_3^{\prime} - |\lambda_5| - \frac{|\lambda_7|}{2} + 2 \sqrt{\lambda_\Phi \lambda_1} \geq 0, \nonumber \\
     \lambda_4 + \lambda_4^{\prime} - \frac{|\lambda_7|}{2} + 2 \sqrt{\lambda_\Phi \lambda_2} \geq 0, \quad \lambda_S + \lambda_S^{\prime} + 2 \sqrt{\lambda_1 \lambda_2} \geq 0.
\end{eqnarray}
The {\it CP}-even neutral component of $\Phi$ acquires a vacuum expectation value (VEV), $\upsilon = 246~\text{GeV}$, triggering spontaneous electroweak symmetry breaking (EWSB), while the discrete $Z_2$ symmetry prevents $S_1$ and $S_2$ from developing VEVs, keeping them inert. 

The term $\lambda_7$ in Eq.~(\ref{eq:higgs_potential}) induces mixing between the singly charged scalars $S_1^+$ and $S_2^+$, giving rise to the physical eigenstates $H_1^\pm$ and $H_2^\pm$ defined as
\begin{equation}
    \begin{pmatrix}
        H_1^\pm \\
        H_2^\pm
    \end{pmatrix} =\underbrace{ \begin{pmatrix}
        \cos\beta && \sin\beta \\
        -\sin\beta && \cos\beta
    \end{pmatrix} }_{R(\beta)} \begin{pmatrix}
        S_1^\pm \\
        S_2^\pm
    \end{pmatrix},
\end{equation}
where $\beta$ represents the mixing angle associated with the diagonalization of the singly charged scalar mass matrix given by
\begin{equation}
M^2_{S_{1,2}^{\pm}} = \frac{1}{2} \begin{pmatrix}
    2 \mu_1^2 + \lambda_3 \upsilon^2 && \lambda_7 \upsilon^2 \\
    \lambda_7 \upsilon^2 && 2\mu_2^2 + \lambda_4 \upsilon^2 + \lambda_4^\prime \upsilon^2
\end{pmatrix}.
\label{MassMixing}
\end{equation}
Thus, the diagonalization is performed when we apply $R(\beta)~M^2_{S_{1,2}^{\pm}}~R(\beta)^T = \text{diag}(m_{H_1^\pm}^2, m_{H_2^\pm}^2)$, leading to the following constraints:
\begin{eqnarray}
    \mu_1^2 + \frac{\lambda_3 \upsilon^2}{2} &=& \cos^2\beta~  m_{H_1^\pm}^2 + \sin^2\beta~m_{H_2^\pm}^2, \quad
     \mu_2^2 + \frac{\lambda_4 \upsilon^2}{2} + \frac{\lambda_4^\prime \upsilon^2}{2} = \sin^2\beta~  m_{H_1^\pm}^2 + \cos^2\beta~m_{H_2^\pm}^2  \nonumber \\
     \frac{\lambda_7 \upsilon^2}{2} &=& \sin\beta \cos\beta~( m_{H_2^\pm}^2 -  m_{H_1^\pm}^2)
     \label{sca_cons1}
\end{eqnarray}
The masses of the physical charged eigenstates $H_1^\pm$ (lighter) and $H_2^\pm$ (heavier) and the rotation angle $\beta$ are given by
\begin{eqnarray}
    m_{H_{1,2}^\pm}^2 &=& \frac{1}{2} \left( \mu_1^2 + \frac{\lambda_3 \upsilon^2}{2} + \mu_2^2 + \frac{\lambda_4 \upsilon^2}{2}\upsilon^2 + \frac{\lambda_4^\prime \upsilon^2}{2} ~\mp~\sqrt{(\mu_2^2 + \frac{\lambda_4 \upsilon^2}{2} + \frac{\lambda_4^\prime \upsilon^2}{2} - \mu_1^2 - \frac{\lambda_3 \upsilon^2}{2})^2 + \lambda_7^2 \upsilon^4} \right) \nonumber \\
    \tan2\beta &=& \frac{\lambda_7 \upsilon ^2}{ \mu_2^2 + \frac{\lambda_4 \upsilon^2}{2} + \frac{\lambda_4^\prime \upsilon^2}{2} - \mu_1^2 - \frac{\lambda_3 \upsilon^2}{2}}
\end{eqnarray}
The remaining physical masses, namely the {\it CP}-even and {\it CP}-odd neutral scalars, as well as the doubly charged scalars, are given by
\begin{eqnarray}
    m_h^2 = 2 \lambda_\Phi \upsilon^2, \quad m_{H, A}^2 = \mu_1^2 + \frac{\lambda_3 \upsilon^2}{2} + \frac{\lambda_3^{\prime} \upsilon^2}{2} \pm \frac{\lambda_5 \upsilon^2}{2}, \quad m^2_{S_2^{\pm\pm}} = \mu_2^2 + \frac{\lambda_4 \upsilon^2}{2}.
    \label{sca_cons2}
\end{eqnarray}
Subsequently, the spectrum after EWSB consists of two {\it CP}-even scalars $\{h, H\}$, one {\it CP}-odd scalar $A$, a pair of singly charged scalars $\{H_1^\pm, H_2^\pm\}$, and a doubly charged scalar $S_2^{\pm\pm}$. Without loss of generality, we take $\lambda_7 \geq 0$ and $\beta \in [0, \pi/2]$. 

The free parameters in the model are taken to be $\lambda_3$, $\lambda_3^\prime$, $\lambda_5$, $\lambda_7$, $\beta$, $g_{i\alpha}$, $\tilde{g}_{i\alpha}$, $m_{H_1^+}$, and $m_{N_{F_i}}$ (which fix $m_{H_2^+}$, $m_{S_2^{\pm \pm}}$ and $m_{H}$). The remaining model parameters have no practical impact on this work, and they are supposed to be appropriately chosen to avoid potential unitary constraints and ensure vacuum stability.
%%%%%%%%%%%%%%%%%%%%%%%%%%%%%%%%%%%%%%%%%%%%%%%%%%%%%%%%%%%
\subsection{Neutrino mass generation}
\begin{figure}[h]
\center{
\includegraphics[width=0.4\columnwidth]{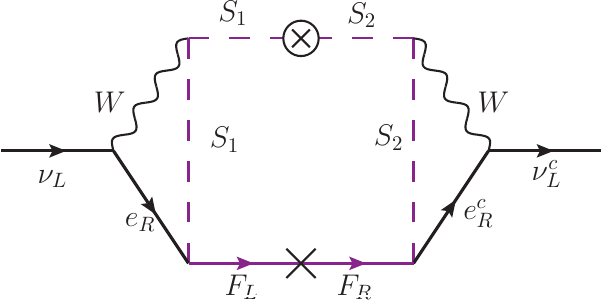}
\hspace{1cm} \includegraphics[width=0.4\columnwidth]{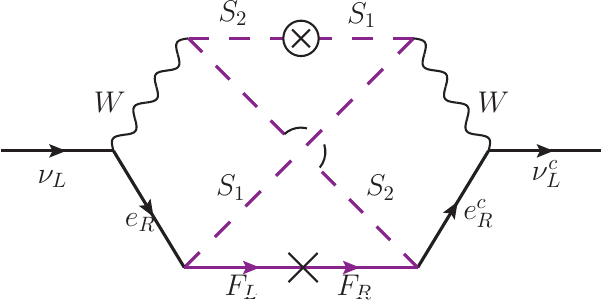}}\\
\center{
\includegraphics[width=0.4\columnwidth]{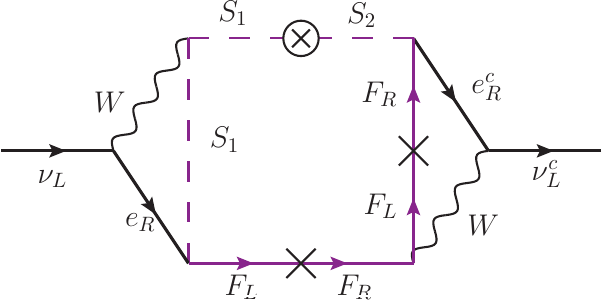}
\hspace{1cm} \includegraphics[width=0.4\columnwidth]{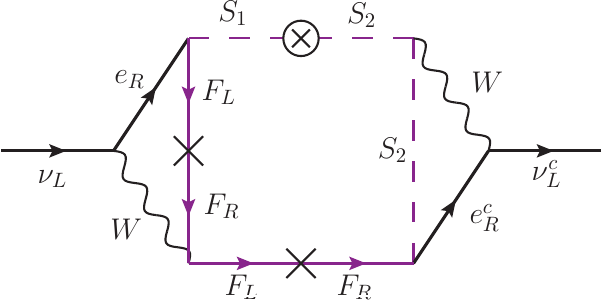}}
\caption{\small Three-loop topologies for generating neutrino masses}
\label{fig_topologies}
\end{figure}

%%%%%%%%%%%%%%%%%%%%%%%%%%%%%%%%%%%%%%%%%%%%%%%%%%%%%%%%%%%%%
The generation of neutrino mass is achieved through a three-loop process encompassing four distinct diagrammatic topologies, as depicted in Fig.~\ref{fig_topologies}. 
Within the framework of the unitary gauge and the mass eigenstate basis, a total of $64$ diagrams are involved. These can be collectively represented by the following expression
\begin{eqnarray}
\label{eq:m_nu}
m^{\nu}_{ab} &=& \frac{\lambda_7 \lambda_5}{(16\pi^2)^3}   \,\times \, (g_a\, \tilde{g}_b + \tilde{g}_a\, g_b )\,  \frac{m_a m_b}{m_W} \,\times  I_{3L}.
\end{eqnarray}
Here, $I_{3L}$ is a dimensionless number and involves a nontrivial three-loop integral that includes all relevant Feynman diagrams. This integral is parametrized by the masses $m_E, m_{H^+_1}, m_{H^+_2}, m_{H^0}, m_{A^0}, m_{S_2^{++}}$ and $\beta$. The computational evaluation of these integrals was performed using the \texttt{Secdec} tool \cite{Borowka:2017idc}. Furthermore, the parameter $\lambda_7$—the mixing  between the charged components of $S_1$ and $S_2$—can be written in terms of the mass difference between the singly charged Higgs bosons, as indicated in Eq. (\ref{sca_cons1}), thus $\lambda_7 v^2 \propto s_\beta c_\beta(m_{H_1^+}^2-m_{H_2^+}^2)$. Similarly, $\lambda_5$ can be expressed as the difference between the squared masses of the neutral {\it CP}-even and {\it CP}-odd scalars, according to Eq. (\ref{sca_cons2}), thus $\lambda_5 \propto m_H^2 - m_A^2$. In order to have a nonzero neutrino mass term, the model imposes that conditions $m_{H^+_1}\neq m_{H^+_2}$ accompanied by $m_H \neq m_A$ are needed.

Another characteristic of the model is the Yukawa structure given by (\ref{eq:lagrangian_yuk}). By examining all topologies in Fig.~\ref{fig_topologies}, the explicit form of the neutrino mass is expressed as $m^{\nu}_{ab} \propto (\tilde{g}_{a} g_{b} + \tilde{g}_{b} g_{a})$, where $a$ and $b$ denote the lepton flavor indices of the neutrino mass matrix elements. This can be written as a product of matrices as follows
\begin{equation}
\label{numass_two}
m^{\nu} \propto \Bigg[
\left(  \begin{array}{ccc}
                  g_{e} & 0 & 0 \\
                  g_{\mu} & 0 & 0 \\
                  g_{\tau} & 0 & 0
                 \end{array} \right) 
                 \,\left(  \begin{array}{ccc}
                  \tilde{g}_{e} & \tilde{g}_{\mu} & \tilde{g}_{\tau} \\
                  0 & 0 & 0 \\
                  0 & 0 & 0 \\
\end{array} \right) +
\left(  \begin{array}{ccc}
                  \tilde{g}_{e} & 0 & 0 \\
                  \tilde{g}_{\mu} & 0 & 0 \\
                  \tilde{g}_{\tau} & 0 & 0
                 \end{array} \right) 
                 \,\left(  \begin{array}{ccc}
                  g_{e} & g_{\mu} & g_{\tau} \\
                  0 & 0 & 0 \\
                  0 & 0 & 0 \\
\end{array} \right) \Bigg]
\end{equation}
The presence of two independent Yukawa matrices, ${g_a, g_b}$ and ${\tilde{g}_a, \tilde{g}_b}$, with symmetrized flavor indices guarantees that the neutrino mass matrix is of rank 2. This is closely analogous to minimal type-I seesaw constructions with two heavy singlet neutrinos, in which one light neutrino remains massless. Consequently, this structure renders the model maximally predictive with regard to the neutrino mass spectrum.

\begin{figure}[H]
\center{\includegraphics[width=0.73 \columnwidth]{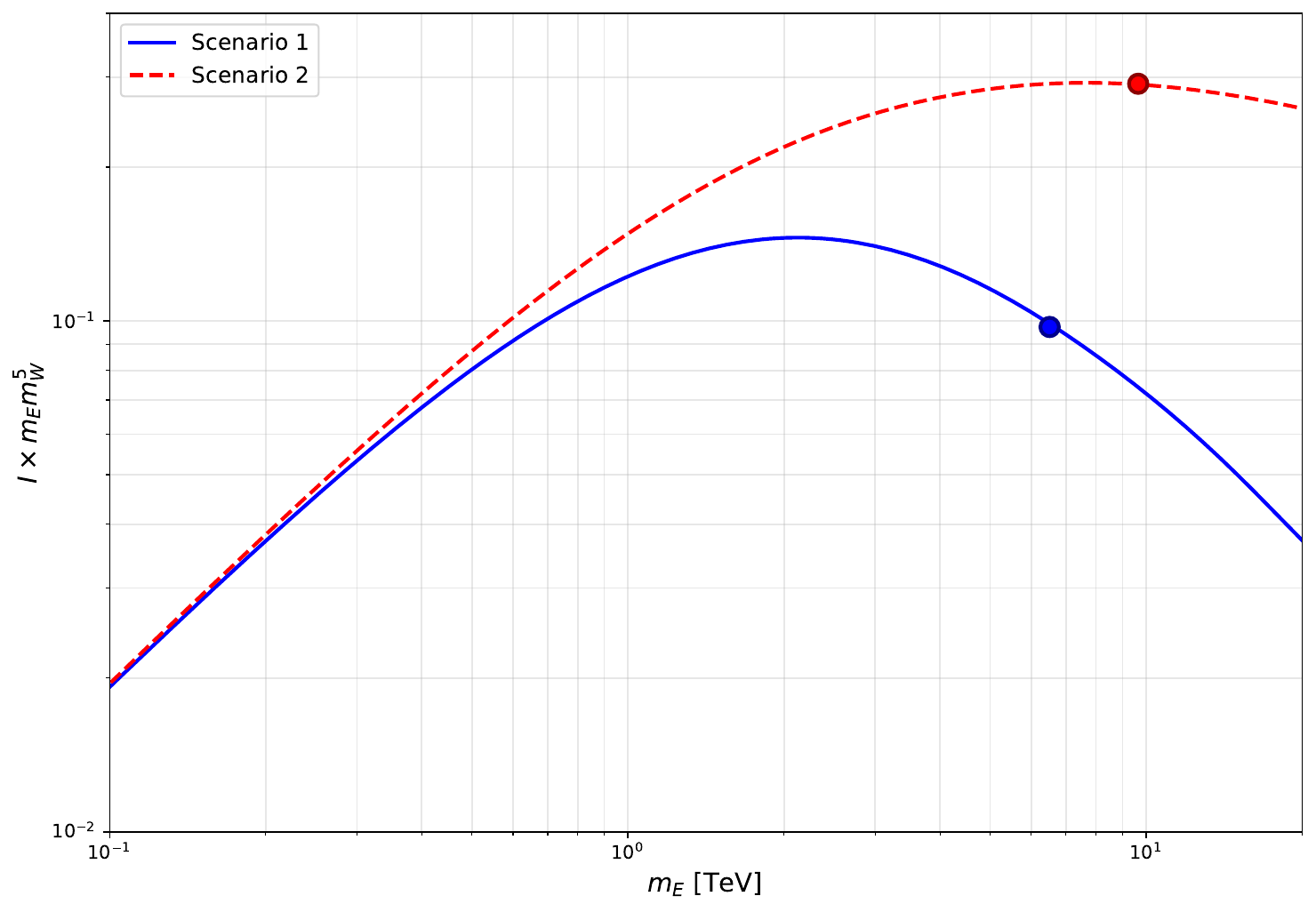}}
\caption{\small Three-loop integral $I_{3L}$ vs VLL mass $m_E$ for two fixed scalar mass spectra. Scenario 1 (2) fixes the scalar parameters to BP1 (BP2) values; continuous curves show $I_{3L}(m_E)$, with highlighted points marking the specific benchmark $m_E$ values.}
\label{integral}
\end{figure}

In Fig.~\ref{integral}, the function $I_{3L}$ is presented as a function of the charged component of the VLL mass $m_E$, illustrating how the three-loop integral $I_{3L}$ depends on the VLL mass $m_E$ for two representative scalar mass spectra that we denote ``Scenario 1" and ``Scenario 2". Each scenario fixes the scalar sector parameters $\{m_H, m_A, m_{H_{1,2}}, m_{S^{++}}, \sin\theta\}$ to the values later used to define benchmark points BP1 and BP2, respectively [Eqs.~(IV.26)--(IV.27)]. Within each scenario, $m_E$ is varied continuously to demonstrate the functional dependence of $I_{3L}$ on this parameter. The discrete highlighted points on each curve correspond exactly to the benchmark values $m_E = 6510.66$~GeV (Scenario 1 $\to$ BP1) and $m_E = 9649.60$~GeV (Scenario 2 $\to$ BP2) employed in the detailed numerical analysis of Sec. IV.

%%%%%%%%%%%%%%%% 
%%%%%%%%%%%%%%%%%%%%%%%%%%%%%%%%%%%%%%%%%%%%%%%%%%%%%%%%%%%%%%%%
\section{Phenomenology}\label{sec:pheno} 
%%%%%%%%%%%%%%%%%%%%%%%%%%%%%%%%%%%%%%%%%%%%%%%%%%%%%%%%%%%%%%%%
\subsection{Dark matter}
%%%%%%%%%%%%%%%%%%%%%%%%%%%%%%%%%%%%%%%%%%%%%%%%%%%%%%%%%%%%%%%%
In this work, the DM candidate corresponds to the lightest $Z_2$-odd particle running in the three-loop diagrams that generate Majorana neutrino masses. In principle, three possibilities arise: a bosonic state, either {\it CP}-even or {\it CP}-odd, or a fermionic state, namely the neutral component of the heavy VLL doublet $\mathcal{F} = (N^0_{L,R}, E^-_{L,R})^T$. However, the fermionic option is strongly constrained. Since $\mathcal{F}$ is an $\mathrm{SU(2)_L}$ doublet with hypercharge $Y=-1/2$, the neutral field $N$ couples to the $Z$ boson through the vertex $\frac{g}{2\cos\theta_W}\, \bar{N}^0 \gamma^\mu N^0\, Z_\mu$, leading to tree-level elastic scattering with nuclei. The corresponding spin-independent (SI) cross section per nucleon is approximately \cite{Cirelli:2005uq}
\begin{equation}
\sigma_{\text{SI}} \simeq \frac{G_F^2\, \mu_N^2}{2\pi} \sim 10^{-38}\,\text{cm}^2 ,
\end{equation}
where $\mu_N$ is the DM nucleon reduced mass. This value, exceeds current experimental upper limits from XENONnT \cite{XENON:2023cxc} and LUX‑ZEPLIN (LZ) \cite{LZ:2024zvo} Collaborations by about ten orders of magnitude. Consequently, a pure Dirac $\mathrm{SU(2)_L}$ doublet fermion with an unsuppressed $Z$ coupling is excluded as a viable DM candidate\footnote{One may opt to extend the model so that the vector-like doublet $\mathcal{F}$ acquires a Majorana mass term, for instance via coupling to a scalar triplet \cite{Bhattacharya:2018fus}, rendering the neutral state Majorana and thereby eliminating the tree-level $Z$-boson mediated elastic scattering.}. Thus, the potential DM candidate  in our model is the lightest scalar running in the three-loop topology, either the {\it CP}-even or the {\it CP}-odd inert state. In principle, taking either particle as the lightest is equivalent, since both states interact identically with the SM fields up to their {\it CP} properties. In our parameter scan, as we will see later, we adopt the same mass range and coupling intervals for both candidates, allowing the numerical analysis to identify regions where either state can serve as the DM particle. Consequently, from the resulting dataset, we select representative benchmark points corresponding to both scenarios.

The parameter space of the present model, in addition to the theoretical constraints discussed in the previous section, is further restricted by the measured value of the DM relic density. Thus, any viable candidate for DM must account for the observed abundance measured  by the Planck Collaboration \cite{Planck:2018vyg}:
\begin{equation}
\Omega_{\rm DM} h^2 = 0.12 \pm 0.001.
\label{oh2}
\end{equation}
In addition, the viable regions of the parameter space are restricted by current sensitivities from DM direct detection experiments, such as XENONnT (2023) \cite{XENON:2023cxc} and LZ (2025) \cite{LZ:2024zvo}, which give the most stringent bound on the DM-nucleus scattering cross section SI $\sigma_{SI} \sim O(10^{-47} - 10^{-48})~\text{cm}^2$.

In the next section, we scan the model's free parameters to identify regions consistent with the relic density measured by Planck. We then determine which portion of the parameter space allowed by the relic-density constraint also satisfies current direct detection bounds. From the subset of points compatible with the relic abundance, direct detection limits, and the theoretical and experimental constraints discussed earlier, we select two benchmark points for a detailed analysis of the neutrino sector and the charged lepton flavor violation (cLFV) process $\mu \rightarrow e \gamma$. Before proceeding, we briefly comment on the general aspects of neutrino phenomenology and cLFV processes.
%%%%%%%%%%%%%%%%%%%%%%%%%%%%%%%%%%%%%%%%%%%%%%%%%%%%%%%%%%%%%%%%
\subsection{Neutrino masses and mixing angles}
%%%%%%%%%%%%%%%%%%%%%%%%%%%%%%%%%%%%%%%%%%%%%%%%%%%%%%%%%%%%%%%%
In this work, we assume that the charged lepton mass matrix is diagonal in the weak basis. As a result, the lepton mixing angles arise solely from neutrino oscillations, and the neutrino mass matrix is diagonalized by the Pontecorvo, Maki, Nakagawa and Sakata (PMNS) matrix, $U_{PMNS}$, which, in the standard parametrization, is given by \cite{ParticleDataGroup:2024cfk}. 
\begin{equation}  
U_{PMNS} =  
\begin{pmatrix}  
c_{12}  c_{13}  & s_{12}  c_{13}  & s_{13}  e^{-i \delta_{CP} } \\  
-s_{12}  c_{23}  - c_{12}  s_{13}  s_{23}  e^{i \delta_{CP} } & c_{12}  c_{23}  - s_{12}  s_{13}  s_{23}  e^{i \delta_{CP} } & c_{13}  s_{23}  \\  
s_{12}  s_{23}  - c_{12}  s_{13}  c_{23}  e^{i \delta_{CP} } & -c_{12}  s_{23}  - s_{12}  s_{13}  c_{23}  e^{i \delta_{CP} } & c_{13}  c_{23}   
\end{pmatrix}.  
\end{equation}
Here, we use the shorthand notation $c_{ij}  \equiv \cos\theta_{ij} $ and $s_{ij}  \equiv \sin\theta_{ij} $ for the lepton mixing angles. The presence of LNV enables the generation of Majorana masses for light neutrinos. As a result, the neutrino mixing matrix must be extended to include additional physical phases associated with the Majorana nature of neutrinos. These are captured by a diagonal phase matrix $P = \text{diag}(1, e^{i\alpha_{21}/2}, e^{i\alpha_{31}/2})$, which multiplies the standard PMNS matrix. 
\begin{table}[h]
\centering
\renewcommand{\arraystretch}{1.2}
\setlength{\tabcolsep}{12pt}
\begin{tabular}{|c|c|c|}
\hline
Parameters & $\mu_i \pm 1\sigma$ & $3\sigma$ ranges \\ \hline
$\Delta m_{21}^2/10^{-5}eV^2$ & $7.49 \pm 0.19$ & $6.92 \rightarrow 8.05$ \\ \hline
$\Delta m_{31}^2/10^{-3}eV^2~\text{(NO)}$ & $2.513_{-0.019}^{+0.021}$ & $2.451 \rightarrow 2.578$ \\ \hline
$\Delta m_{32}^2/10^{-3}eV^2~\text{(IO)}$ & $-2.484 \pm 0.020$ & $-2.547  \rightarrow -2.421$ \\ \hline
$\sin^2\theta_{12} $ & $0.308_{-0.011}^{+0.012}$ & $0.275 \rightarrow 0.345$ \\ \hline
$\sin^2\theta_{23}~\text{(NO)} $ & $0.470_{-0.013}^{+0.017}$ & $0.435 \rightarrow 0.585 $ \\ \hline
$\sin^2\theta_{23}~\text{(IO)} $ & $0.550_{-0.015}^{+0.012}$ & $0.440 \rightarrow 0.584 $ \\ \hline
$\sin^2\theta_{13}~\text{(NO)}$ & $0.02215_{-0.00058}^{+0.00056}$ & $0.02030 \rightarrow 0.02388$ \\ \hline
$\sin^2\theta_{13}~\text{(IO)}$ & $0.02231 \pm 0.00056$ & $0.02060 \rightarrow 0.02409$ \\ \hline
$\delta_{CP} /^\circ~\text{(NO)}$ & $212_{-41}^{+26}$ & $124 \rightarrow 364$ \\ \hline
$\delta_{CP} /^\circ~\text{(IO)}$ & $274_{-25}^{+22}$ & $201 \rightarrow 335$ \\ \hline
$m_e / m_\mu$ & $0.004737 \pm 4.737 \times 10^{-6}$ & $0.0047228 \rightarrow 0.0047512$ \\ \hline
$m_\mu / m_\tau$ & $0.05882 \pm 5.882 \times 10^{-5}$ & $0.0586435 \rightarrow 0.0589965$ \\
\hline
\end{tabular}
\caption{The global best fit, $1\sigma$ and $3\sigma$ ranges for the lepton mixing angles, neutrino mass squared differences, and leptonic {\it CP}-violating phase from the latest global fit NuFIT v6.0, including SK atmospheric data \cite{Esteban:2024eli}. The central values of the charged lepton mass ratios are taken from \cite{Xing:2007fb}. We assign relative uncertainties of $0.1\%$ (1$\sigma$) and $0.3\%$ (3$\sigma$) for the purpose of parameter scanning and fitting.}
\label{data}
\end{table}
The full neutrino mixing matrix is therefore given by $U_\nu = U_{\text{PMNS}} P$, where $U_{\text{PMNS}}$ governs the flavor mixing, and $P$ encodes the Majorana {\it CP}-violating phases. Information on these phases could potentially be provided by neutrinoless double beta decay ($0\nu\beta\beta$) experiments. The decay amplitude of this process is proportional to the effective Majorana neutrino mass $|m_{\beta\beta}|$ defined as $|m_{\beta\beta}| = |\sum_i U_{\nu_{e_i}}^2 m_i|$. In addition to $|m_{\beta\beta}|$, two other key observables provide insights into the absolute neutrino mass scale. The first is the sum of the three neutrino masses, $\sum m_i = m_1 + m_2 + m_3$, which is constrained by cosmological data. The second is the effective electron antineutrino mass $m_\beta$, measurable through precise analysis of the electron energy spectrum in low-energy beta decay experiments. This effective mass is defined as $m_\beta = (\sum_i |U_{\nu_{e_i}}|^2 m_i^2)^{1/2}$. In this study, we numerically analyze both normal and inverted neutrino mass orderings (NO and IO). Moreover, in order to facilitate our numerical fit of the neutrino observables, we rewrite Eq.~(\ref{eq:m_nu}) in the following manner
\begin{equation}
    m_{ab}^\nu = \frac{\lambda_5 \lambda_7}{(16 \pi^2)^3} \frac{m_\mu^2}{m_W} I_{3L} \begin{pmatrix}
        2 \frac{m_{e}^2}{m_\mu^2} g_e \tilde{g}_e & \frac{m_e}{m_\mu} (g_e \tilde{g}_\mu + \tilde{g}_e g_\mu) & \frac{m_e}{m_\mu}\frac{m_\tau}{m_\mu} (g_e \tilde{g}_\tau + \tilde{g}_e g_\tau) \\
        \frac{m_e}{m_\mu} (g_e \tilde{g}_\mu + \tilde{g}_e g_\mu) & 2  g_\mu \tilde{g}_\mu &  \frac{m_\tau}{m_\mu} (g_\mu \tilde{g}_\tau + \tilde{g_\mu g_\tau}) \\
        \frac{m_e}{m_\mu}\frac{m_\tau}{m_\mu} (g_e \tilde{g}_\tau + \tilde{g}_e g_\tau) & \frac{m_\tau}{m_\mu} (g_\mu \tilde{g}_\tau + \tilde{g_\mu g_\tau}) & 2 \frac{m_\tau^2}{m_\mu^2} g_\tau \tilde{g}_\tau
    \end{pmatrix},
    \label{nu-mass}
\end{equation}
where the muon mass and $W$ boson mass are fixed from Ref. \cite{ParticleDataGroup:2024cfk}. The numerical values of the oscillation parameters and the charged lepton mass ratios used in our numerical scan can be found in Table~\ref{data}.
%%%%%%%%%%%%%%%%%%%%%%%%%%%%%%%%%%%%%%%%%%%%%%%%%%%%%%%%%%%%%%%%
\subsection{Charged lepton flavor violation processes}
%%%%%%%%%%%%%%%%%%%%%%%%%%%%%%%%%%%%%%%%%%%%%%%%%%%%%%%%%%%%%%%%
Charged lepton flavor violation processes provide important experimental constraints mainly through the decay $\mu \to e \gamma$.  In general, the decay width of the radiative cLFV process $l_j \to l_i \gamma$ can be expressed as
\begin{equation}
    \Gamma(l_j \to l_i \gamma) \;=\; \frac{1}{16\pi\,m_{l_j}^3}\,(m_{l_j}^2-m_{l_i}^2)^3\big(|\sigma^{ij}_L|^2+|\sigma^{ij}_R|^2\big),
\end{equation}
where $\sigma^{ij}_L$ and $\sigma^{ij}_R$ are the form factors from beyond the SM loop contribution (see \cite{Lavoura:2003xp} for a general case).
Using the tree-level width $\Gamma(l_j \to l_i \bar{\nu}_j \nu_i)=G_F^2 m_{l_j}^5/(192\pi^3)$, the branching ratio can be expressed as
\begin{equation}
    \mathrm{BR}(l_j \to l_i \gamma)
    \;=\;
    \frac{\Gamma(l_j \to l_i \gamma)}{\Gamma(l_j \to l_i \bar{\nu}_j \nu_i)}
    \times
    \mathrm{BR}(l_j \to l_i \bar{\nu}_j \nu_i).
\end{equation}
In the limit $m_{l_i}\ll m_{l_j}$ (which we adopt in the following), this simplifies to
\begin{equation}
\label{BR:general}
    \mathrm{BR}(l_j \to l_i \gamma)
    \;=\;
    \frac{12\pi^2}{G_F^2 m_{l_j}^2}\big(|\sigma^{ij}_L|^2+|\sigma^{ij}_R|^2\big)\,
    \times \mathrm{BR}(l_j \to l_i \bar{\nu}_j \nu_i).
\end{equation}
Based on the results given in Ref.~\cite{Lavoura:2003xp}, the form factors can be written as
\[
\sigma^{ij}_L = e\, m_{l_j}\, \xi \, A_L^{ij},\qquad
\sigma^{ij}_R = e\, m_{l_i}\, \xi \, A_R^{ij},
\]
where $\xi=g_{i}^* g_{j}$ and $\xi=\tilde{g}_{i}^* \tilde{g}_{j}$ which depend on the particular diagram involved. When the SM charged-lepton masses are negligible compared to the masses of particles running in the loop, we have $A_L^{ij}=A_R^{ij}\equiv A^{ij}$, thus Eq.~(\ref{BR:general}) becomes
\begin{equation}
    \mathrm{BR}(l_j \to l_i \gamma)
    \;=\;
    \frac{48\pi^3 \alpha}{G_F^2}\left(1+\frac{m_{l_i}^2}{m_{l_j}^2}\right)|A^{ij}|^2\,
    \times
    \mathrm{BR}(l_j \to l_i \bar{\nu}_j \nu_i),
    \label{BRemg}
\end{equation}
which, under $m_{l_i}\ll m_{l_j}$, reduces to the expression commonly used with the factor $(1+m_{l_i}^2/m_{l_j}^2)\simeq 1$ (or equivalently, $\sigma^{ij}_R \ll \sigma^{ij}_L$).

The one-loop photon-emission amplitude induced by the charged VLL $E^-$ and by singly and doubly charged scalars can be written as
\begin{align}
{16\pi^2}\, m_N^2\, A^{ij}
&= g_i^* g_j\big[c_\beta^2 I_S(t_1)+s_\beta^2 I_S(t_2)-\tfrac{1}{2}I_S(t_H)-\tfrac{1}{2}I_S(t_A)\big] \nonumber\\[4pt]
&\quad + \tilde{g}_i^* \tilde{g}_j\big[s_\beta^2 I_S(t_1)+c_\beta^2 I_S(t_2)+2 I_S(t_{S^{++}})- I_F(t_{S^{++}})\big],
\label{eq:Aij}
\end{align}
where
\[
t_1=\frac{m_N^2}{m_{H_1^+}^2},\quad
t_2=\frac{m_N^2}{m_{H_2^+}^2},\quad
t_H=\frac{m_N^2}{m_H^2},\quad
t_A=\frac{m_N^2}{m_A^2},\quad
t_{S^{++}}=\frac{m_N^2}{m_{S^{++}}^2},
\]
while the scalar and fermion loop functions are given by\footnote{The expressions used here have been kept from Eqs. (40) and (41) in Ref. \cite{Lavoura:2003xp} multiplied by $t$.}
\begin{eqnarray}
    I_F(t) = \frac{t^4 - 6t^3 + 3t^2 + 2t + 6t^2 \ln t}{12(t-1)^4}, \quad
I_S(t) = \frac{2t^4 + 3t^3 - 6t^2 + t - 6t^3 \ln t}{12(t-1)^4}.
\end{eqnarray}
%%%%%%%%%%%%%%%%%%%%%%%%%%%%%%%%%%%%%%%%%%%%%%%%%%%%%%%%%%%%%%%%%%%%
\section{Numerical results}
\label{sec:numres} 
%%%%%%%%%%%%%%%%%%%%%%%%%%%%%%%%%%%%%%%%%%%%%%%%%%%%%%%%%%%%%%%%%%%%
In this section, we present a numerical analysis of the DM sector and the lepton flavor structure predictions within the model introduced above.
%%%%%%%%%%%%%%%%%%%%%%%%%%%%%%%%%%%%%%%%%%%%%%%%%%%%%%%%%%%%%%%%%%%
\subsection{DM relic density and direct detection}
%%%%%%%%%%%%%%%%%%%%%%%%%%%%%%%%%%%%%%%%%%%%%%%%%%%%%%%%%%%%%%%%%%%
The relevant parameters for the computation of the thermal component to the DM relic abundance in our model are the masses of inert states $m_H$, $m_A$, $m_{H_{1,2}^\pm}$, $m_{S_2^{\pm\pm}}$, and their couplings to the Higgs bosons, namely $\lambda_3$, $\lambda_3^\prime$, $\lambda_5$ and $\lambda_7$, the VLL masses $m_E = m_N$, and the new Yukawa couplings $g_l$ and $\tilde{g}_l$ with $l=e,\mu,\tau$. As we discussed above, the couplings $\lambda_5$ and $\lambda_7$ are essential to neutrino mass generation and are proportional to the mass differences between the neutral and charged inert states, respectively. To accurately capture the key features of the DM phenomenology, the model is implemented in \texttt{FeynRules} \cite{Alloul:2013bka}, which is used to generate the corresponding \texttt{CalcHEP} model files \cite{Belyaev:2012qa}. These files are then utilized within \texttt{micrOMEGAs} \cite{Belanger:2013oya,Alguero:2023zol} to perform numerical analyses and compute the dark matter relic abundance. Our numerical scan is performed by varying the input parameters as follows:
\begin{eqnarray}
    m_H, m_A, m_{H_{1,2}^\pm} &\in& [100 - 2000] \text{GeV}, \quad m_{S_2^{\pm\pm}} \in [800 - 2000] \text{GeV}, \nonumber \\ m_E = m_N &\in& [900 - 10000] \text{GeV}, \quad \sin\beta \in [0-1]
\end{eqnarray}
while the coupling constants are required to satisfy the perturbative regime in Eq. (\ref{perturbativity}).

After identifying the region of the parameter space that is consistent with the observed DM relic abundance, we next confront these scenarios with bounds from direct detection experiments. In direct detection, DM may scatter elastically off nuclei in terrestrial detectors. In this framework, the dominant contribution comes from the SI scattering mediated by the Higgs boson at tree level. The corresponding cross section reads as
\begin{equation}
\sigma_{SI} = \frac{\lambda_{33'5}^2\, f_n^2}{4\pi}\,\frac{m_n^4}{m_h^4\,(m_n + m_{DM})^{2}},
\end{equation}
where $m_n$ is the nucleon mass, $m_h = 125~\text{GeV}$ is the SM-like Higgs boson mass \cite{ATLAS:2012yve,CMS:2013btf}, and $f_n$ denotes the nucleon form factor for the scalar interaction \cite{Mambrini:2011ik,Kanemura:2010sh,Alarcon:2011zs,Alarcon:2012nr}.
\begin{figure}[!ht]
\centering
\includegraphics[width=0.45 \columnwidth]{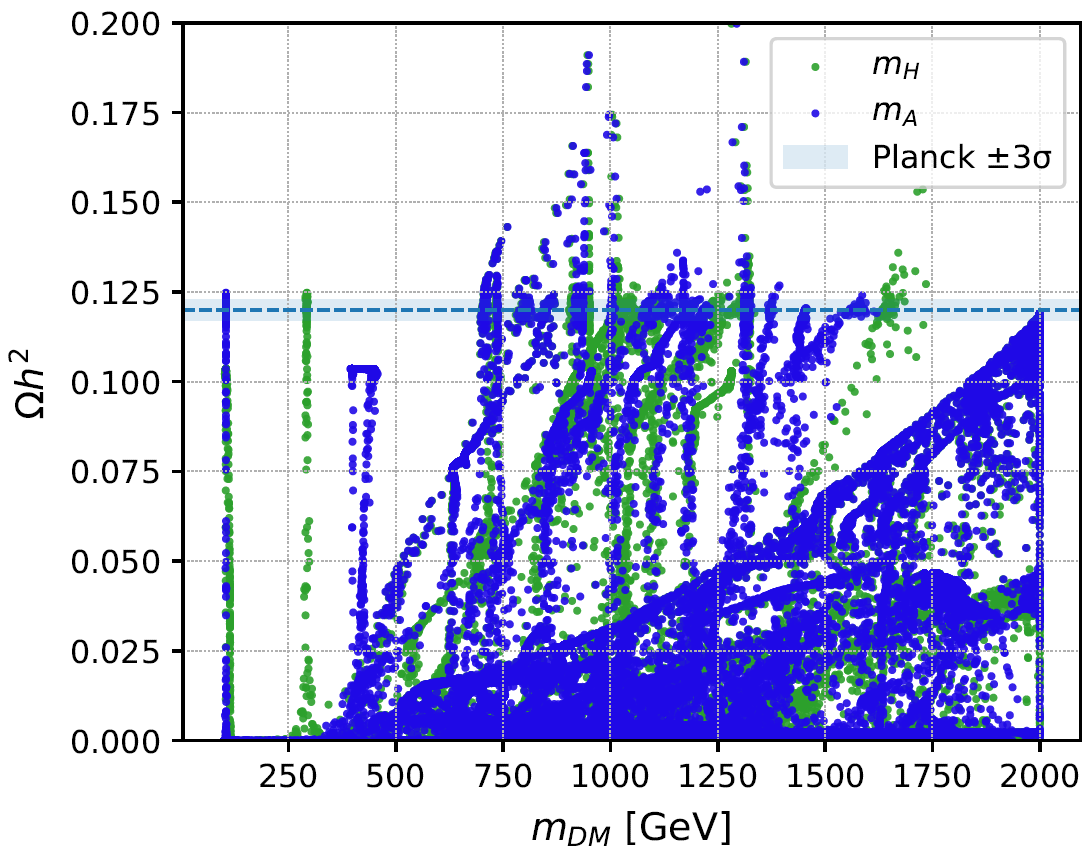}
\hspace{0.3cm}
\includegraphics[width=0.45 \columnwidth]{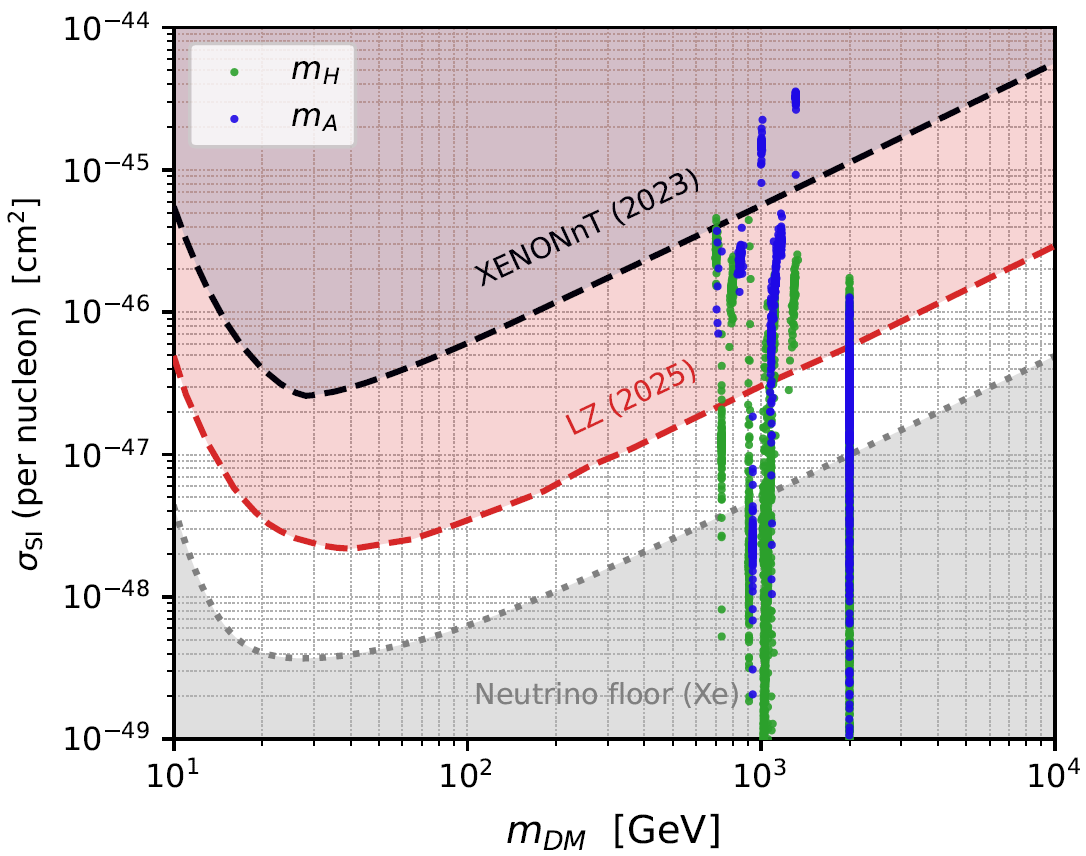}
\caption{Left: dark matter abundance for the {\it CP}-even scalar $H^0$ (green) and {\it CP}-odd scalar $A^0$ (blue). Right: spin-independent dark matter-nucleon cross section as a function of DM mass for $H^0$ (green) and $A^0$ (blue). Also shown are current bounds from XENONnT (dashed black) and LZ (dashed red), along with the neutrino floor for a xenon target (shaded gray). The displayed points correspond to phenomenologically viable solutions from our numerical scan and illustrate where consistent regions occur; they are not intended to provide a statistically weighted or exhaustive coverage of the full parameter space.}
\label{figs}
\end{figure}
In analogy with the inert doublet model \cite{Deshpande:1977rw,Barbieri:2006dq}, the effective Higgs-DM coupling $\lambda_{33'5}$ is given by $\lambda_{33'5} = \lambda_L \equiv \lambda_3 + \lambda_3' + \lambda_5$ for $m_{DM} \equiv m_H$ and $\lambda_{33'5} = \lambda_A \equiv \lambda_3 + \lambda_3' - \lambda_5$ for $m_{DM} \equiv m_A$.

In the left panel of Fig.~\ref{figs}, we present the relic density as a function of the mass of the scalar DM candidates, $m_H$ (green) and $m_A$ (blue). The horizontal band denotes the $3\sigma$ range for cold DM as measured by the Planck satellite, see Eq.~(\ref{oh2}). For both candidates, the viable parameter space predominantly corresponds to heavy DM with masses above $700~\text{GeV}$, while most of the points outside this experimental band result in an underabundance of DM. The right panel of Fig.~\ref{figs} displays the SI DM-nucleon scattering cross section as a function of the DM mass, considering only the parameter points consistent with the observed relic density from the left panel. Current exclusion limits from the XENONnT (dashed black line) and LZ (dashed red line) Collaborations are shown, together with the projected neutrino floor (gray shaded region) from Ref.~\cite{OHare:2021utq}. For DM masses below $1~\text{TeV}$, the dominant allowed candidate is the {\it CP}-even scalar $H^0$. Near $1~\text{TeV}$, both $H^0$ and $A^0$ candidates can satisfy the constraints of the relic density and direct detection. Parameter points beneath the neutrino floor are expected to remain undetectable in direct detection experiments. Conversely, scenarios with small Higgs-portal couplings that place them above the neutrino floor remain potentially accessible to collider searches. The constraints on the Higgs-DM coupling are shown in Fig.~4 for the scalar DM candidate $H^0$, where the relic density is plotted as a function of the Higgs-portal coupling $\lambda_L$. 
As in inert doublet models, the scalar potential is invariant under the interchange $H^0 \leftrightarrow A^0$ accompanied by $\lambda_L \leftrightarrow \lambda_A$ (equivalently, a sign change of $\lambda_5$). This symmetry leaves the relevant annihilation and coannihilation cross sections unchanged, implying identical relic density predictions for the pseudoscalar candidate $A^0$. For this reason, we display only one representative panel.
\begin{figure}[!ht]
\centering
\includegraphics[width=0.48 \columnwidth]{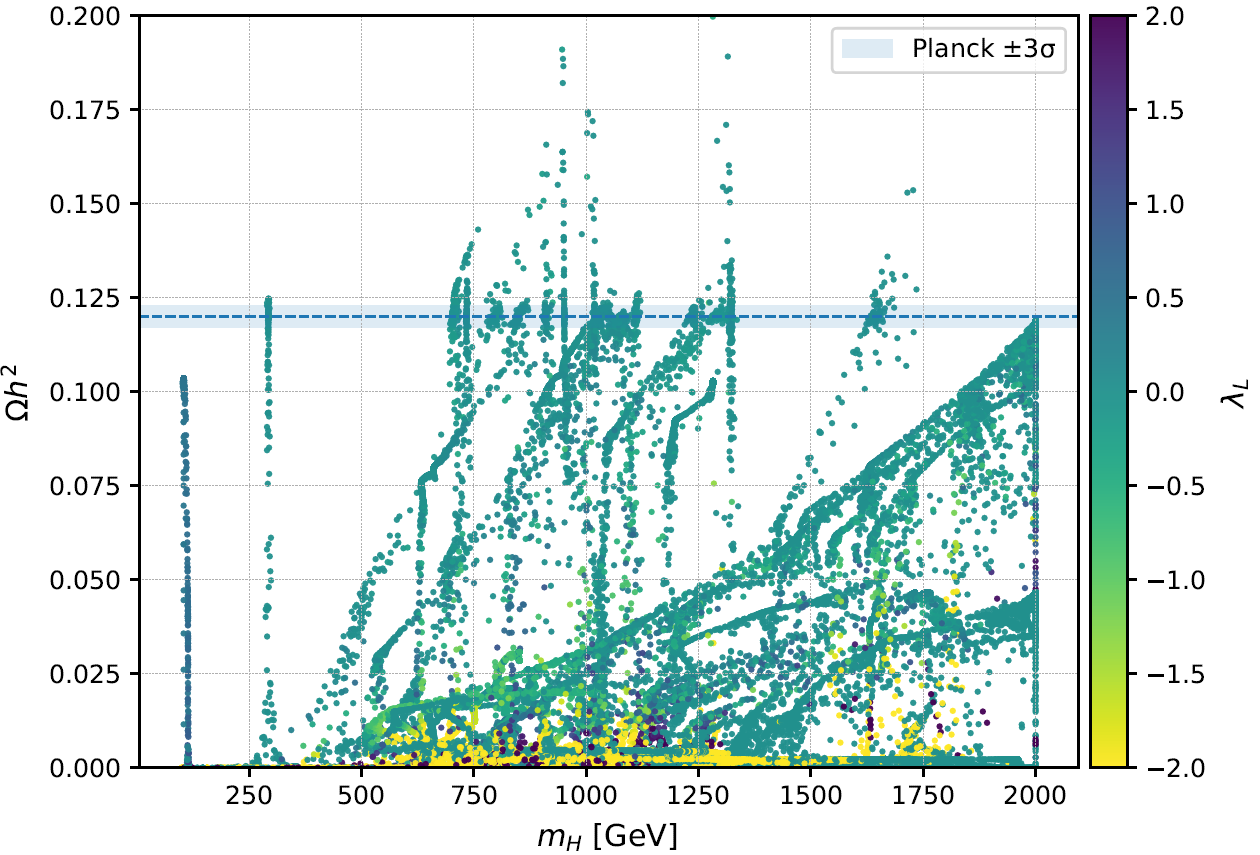}
\caption{Relic density as a function of the DM mass $m_{H^0}$. The color scale indicates the value of the Higgs-portal coupling $\lambda_L$.}
\label{figs2}
\end{figure}
Then, we select from the allowed parameter space two representative benchmark points to carry out the neutrino phenomenology analysis presented in the next section. These benchmark points correspond to the highlighted points in Fig. \ref{integral}: BP1 realizes Scenario 1 at $m_E = 6510.66$~GeV, while BP2 realizes Scenario 2 at $m_E = 9649.60$~GeV.
The chosen benchmark are\footnote{The relatively large quartic couplings in these benchmark points satisfy our tree-level consistency conditions and reflect the particular scalar mass spectra chosen to realize heavy DM with the correct relic abundance and viable mass splittings. They are not enforced by a single phenomenological constraint, and other viable parameter points with smaller quartic values exist. A full renormalization-group analysis, which would require specifying the ultraviolet scale up to which the model is assumed to remain perturbative, is beyond the scope of the present work.}
\begin{eqnarray}
    BP 1 :~~m_H &=& 842.52~\text{GeV}, \quad m_A = 852.94~\text{GeV}, \quad m_{H_1^\pm} = 1165.18~\text{GeV}, \nonumber \\ 
    m_{H_2^\pm} &=& 1100.95~\text{GeV}, \quad m_{S_2^{\pm \pm}} = 1442.94~\text{GeV}, \quad m_E = m_N = 6510.66~\text{GeV} \nonumber \\
    \lambda_3 &=& -9.61, \quad \lambda_3^\prime = 9.86, \quad \sin\beta = 0.139
    \label{bp1}
\end{eqnarray}
where the DM candidate corresponds to the \textit{CP}-even scalar $H$, and
\begin{eqnarray}
    BP 2 :~~m_H &=& 1180.99~\text{GeV}, \quad m_A = 997.83~\text{GeV}, \quad m_{H_1^\pm} = 1150.39~\text{GeV}, \nonumber \\ 
    m_{H_2^\pm} &=& 1147.41~\text{GeV}, \quad m_{S_2^{\pm \pm}} = 1697.01~\text{GeV}, \quad m_E = m_N = 9649.60~\text{GeV} \nonumber \\
    \lambda_3 &=& 8.91, \quad \lambda_3^\prime = -2.29, \quad \sin\beta = 0.757
    \label{bp2}
\end{eqnarray}
where the DM candidate is the \textit{CP}-odd scalar $A$. For BP1, the masses of the singly and doubly charged $Z_2$-odd states lie above the DM mass by approximately $+258$, $+323$, and $+600~\mathrm{GeV}$ for $H_2^\pm$, $H_1^\pm$, and $S_2^{\pm\pm}$, respectively, while the neutral partner is only $\Delta_0 \equiv m_A-m_H = 10.42~\text{GeV}$ heavier. The charged sector mixing is small, $\sin\beta=0.139$, which fixes $\lambda_7=-0.660$ from the $(H_2^\pm,H_1^\pm)$ splitting, while the neutral splitting $\Delta_0$ fixes $\lambda_5=-0.291$. At freeze-out ($x_f=25.5$) the thermal bath is predominantly neutral, with population fractions $H:57.32\%$ and $A:42.60\%$, and charged species $\ll 1\%$. Correspondingly, the effective (co)annihilation cross section at freeze-out is dominated by electroweak gauge self-annihilations in the neutral sector: $HH\!\to W^+W^-$ and $HH\!\to ZZ$ contribute about $43\%$ and $28\%$, respectively, while $AA\!\to hh$, $AA\!\to W^+W^-$, $AA\!\to ZZ$, and $AA\!\to t\bar t$ contribute about $12\%$, $8\%$, $5\%$, and $1\%$ (the remainder coming from channels below the percentage level). The resulting relic abundance is $\Omega h^2=0.120$, and VLLs at $m_{\rm VLL}\!\sim\!6.5~\mathrm{TeV}$ are thermally decoupled and irrelevant for freeze-out. For BP2, the neutral mass splitting is $\Delta_0 \equiv m_H-m_A = 183~\mathrm{GeV}$, which suppresses $AH$ coannihilation, while the singly charged states are moderately heavier by $150$-$153~\mathrm{GeV}$. The mixing is sizable, $\sin\beta=0.757$, corresponding to $\lambda_7=-0.112$, and the neutral splitting gives $\lambda_5=+6.58$. At freeze-out ($x_f=25.3$) the thermal composition features $A$ at $89.3\%$, with few-percentage populations for $H_2^\pm$ ($\sim 5.0\%$) and $H_1^\pm$ ($\sim 4.6\%$) and about $1.1\%$ for $H$; nevertheless, the effective (co)annihilation cross section is almost entirely saturated by gauge self-annihilations of the DM, with $AA\!\to W^+W^-$ and $AA\!\to ZZ$ contributing about $53\%$ and $46\%$, respectively (each remaining coannihilation channel contributing below the percentage level), yielding $\Omega h^2=0.120$, while VLLs near $9.6~\mathrm{TeV}$ remain thermally irrelevant.
%%%%%%%%%%%%%%%%%%%%%%%%%%%%%%%%%%%%%%%%%%%%%%%%%%%%%%%%%%%%%%%%%%%
\subsection{Neutrino phenomenology and cLFV}
%%%%%%%%%%%%%%%%%%%%%%%%%%%%%%%%%%%%%%%%%%%%%%%%%%%%%%%%%%%%%%%%%%%
Here, we present a numerical analysis of the lepton flavor structure predictions within the model introduced above. The neutrino mass matrix, as defined in Eq.~(\ref{nu-mass}), depends on six complex free Yukawa couplings $(g_e, g_\mu, g_\tau, \tilde{g}_e, \tilde{g}_\mu, \tilde{g}_\tau)$ and the charged lepton mass ratios $m_e/m_\mu$ and $m_\tau/m_\mu$. The overall mass scale of the neutrino matrix is controlled by a prefactor that incorporates the standard factor associated with three loops $1/(16 \pi^2)^3$, the scalar couplings $\lambda_5$ and $\lambda_7$, the squared muon mass $m_\mu^2$, the inverse of the $W$ boson mass, and the three-loop integral factor $I_{3L}$. Together, these elements determine the structure of the neutrino masses, lepton mixing angles, and {\it CP}-violating phases. 

The neutrino oscillation parameters, together with the observables discussed in the previous section—$m_{\beta\beta}$, $\sum m_i$, and $m_\beta$—as well as the unknown Majorana phases, provide testable predictions that allow us to evaluate the validity of our model. To assess how well the model agrees with the experimental data on the masses and mixing angles, we performed a $\chi^2$ analysis. The function $\chi^2$ is defined as
\begin{equation}
    \chi^2 = \sum_i \left( \frac{Q_i(x) - \mu_i}{\sigma_i} \right)^2
\end{equation}
Here, $\mu_i$ and $\sigma_i$ denote the experimental central values and $1\sigma$ uncertainties of the eight flavor observables listed in Table~\ref{data}, while $Q_i(x)$ represents the model predictions for these observables, evaluated at the point $x$ in the parameter space. To extract the neutrino masses and mixing angles, we perform a singular value decomposition of the neutrino mass matrix and compute the corresponding function $\chi^2$. All dimensionless model parameters are treated as independent variables and are varied in magnitude according to the bounds specified in Eq.~(\ref{perturbativity}). The total $\chi^2$ is minimized numerically using the \texttt{FlavorPy} package~\cite{FlavorPy}, which employs the \texttt{lmfit} \cite{newville2016lmfit} algorithm for optimization. To estimate the uncertainty and explore the parameter space around the best-fit point, a Markov Chain Monte Carlo scan is also performed. A model point is considered phenomenologically viable if all eight observables fall within their $1\sigma$ experimental ranges. For our analysis, we retain only those points with $\chi^2 < 1$. 
\begin{table}[H]
\centering
\begin{tabular}{c||c|c|c|c|}
Parameters & \textbf{BP 1 (NO)} & \textbf{BP 1 (IO)} & \textbf{BP 2 (NO)} & \textbf{BP 2 (IO)}  \\
\hline
$g_e$ & $2.727489 - 1.174969~i$ & $-0.000928 - 0.000870~i$ & $0.0010009  + 0.003165~i$ & $-0.000082 + 0.000130~i$ \\
$\tilde{g}_e$ & $3.480377 - 2.210725~i$ & $0.000077 + 0.000076~i$ & $-0.007872 + 0.005919~i$ & $-0.001268 + 0.002984~i$ \\
$g_\mu$ & $0.005675 + 0.000104~i$ & $1.285081 - 0.260278~i$ & $-0.008622 + 0.026778~i$ & $-0.000756 - 0.003972~i$ \\
$\tilde{g}_\mu$ & $-0.002796 - 0.000040~i$ & $1.718869 - 1.981331~i$ & $-0.210614 - 0.188422~i$ & $0.032542 - 0.035362~i$ \\
$g_\tau$ & $0.003060 + 0.166340~i$ & $0.017642 + 0.007043~i$ & $3.453573 + 3.514446~i$ & $-2.040803 - 2.998995~i$ \\
$\tilde{g}_\tau$ & $-0.014825 - 0.017536~i$ & $-0.002117 + 0.001909~i$ & $3.539999 - 3.538779~i$ & $0.369664 + 3.440837~i$ \\
$I_{3L}$ & $0.097525$ & $0.097525$ & $0.29119$ & $0.29119$  \\
\hline
$m_e/m_\mu$ & $0.004737$ & $0.004737$ & $0.004737$ & $0.004737$ \\
$m_\mu/m_\tau$ & $0.058819$ & $0.058819$ & $0.058820$ & $0.058819$  \\
$m_1~(\text{meV})$ & $0$ & $49.087$ & $0$ & $49.084$ \\
$m_2~(\text{meV})$ & $8.653$ & $49.844$ & $8.653$ & $49.841$ \\
$m_3~(\text{meV})$ & $50.133$ & $0$ & $50.137$ & $0$ \\
$\sin^2\theta_{12}$ & $0.3051$ & $0.3099$ & $0.3051$ & $0.3100$  \\
$\sin^2\theta_{13}$ & $0.02219$ & $0.02235$ & $0.02219$ & $0.02231$  \\
$\sin^2\theta_{23}$ & $0.47000$ & $0.54869$ & $0.47045$ & $0.55111$  \\
$\delta_{CP} /\pi$ & $1.166668$ & $1.494017$ & $1.166457$ & $1.511201$  \\
$\alpha_{21}/\pi$ & $1.407370$ & $1.731936$ & $0.890503$ & $0.696671$ \\
$m_\beta~(\text{meV})$ & $8.932$ & $49.794$ & $8.932$ & $49.792$ \\
$m_{\beta\beta}~(\text{meV})$ & $3.681$ & $47.947$ & $3.224$ & $47.525$  \\
$\sum{m_i}~(\text{meV})$ & $58.787$ & $98.932$ & $58.790$ & $98.926$ \\
\hline
$\chi^{2}$ & $0.01122$ & $0.0575$ & $0.01071$ & $0.00485$
\end{tabular}
\caption{Best-fit values of the new Yukawa couplings and the resulting predictions for lepton masses, mixing angles, and {\it CP} phases for BP 1 and BP 2 from the DM analysis, shown for both neutrino mass orderings.}
\label{results}
\end{table}
As previously mentioned, we work in the diagonal charged lepton basis, where the charged lepton mass matrix is given by $M_e = \text{diag}(p_1, p_2, p_3)$. Instead of fixing $p_i$ to the physical lepton masses, we impose the experimental mass ratios $m_e/m_\mu = 0.004737$ and $m_\mu/m_\tau = 0.05882$, as listed in Table~\ref{data}. This is justified by the structure of the neutrino mass matrix in our model, where the charged lepton mass ratios appear due to loop-level mass insertions; see Eq.~(\ref{nu-mass}). These ratios are therefore taken as input parameters in the \texttt{FlavorPy} implementation, enabling consistent numerical scans that maintain the charged lepton mass hierarchy without fixing their absolute values. 
\begin{figure}[h!]
\centering
\includegraphics[width=0.98 \columnwidth]{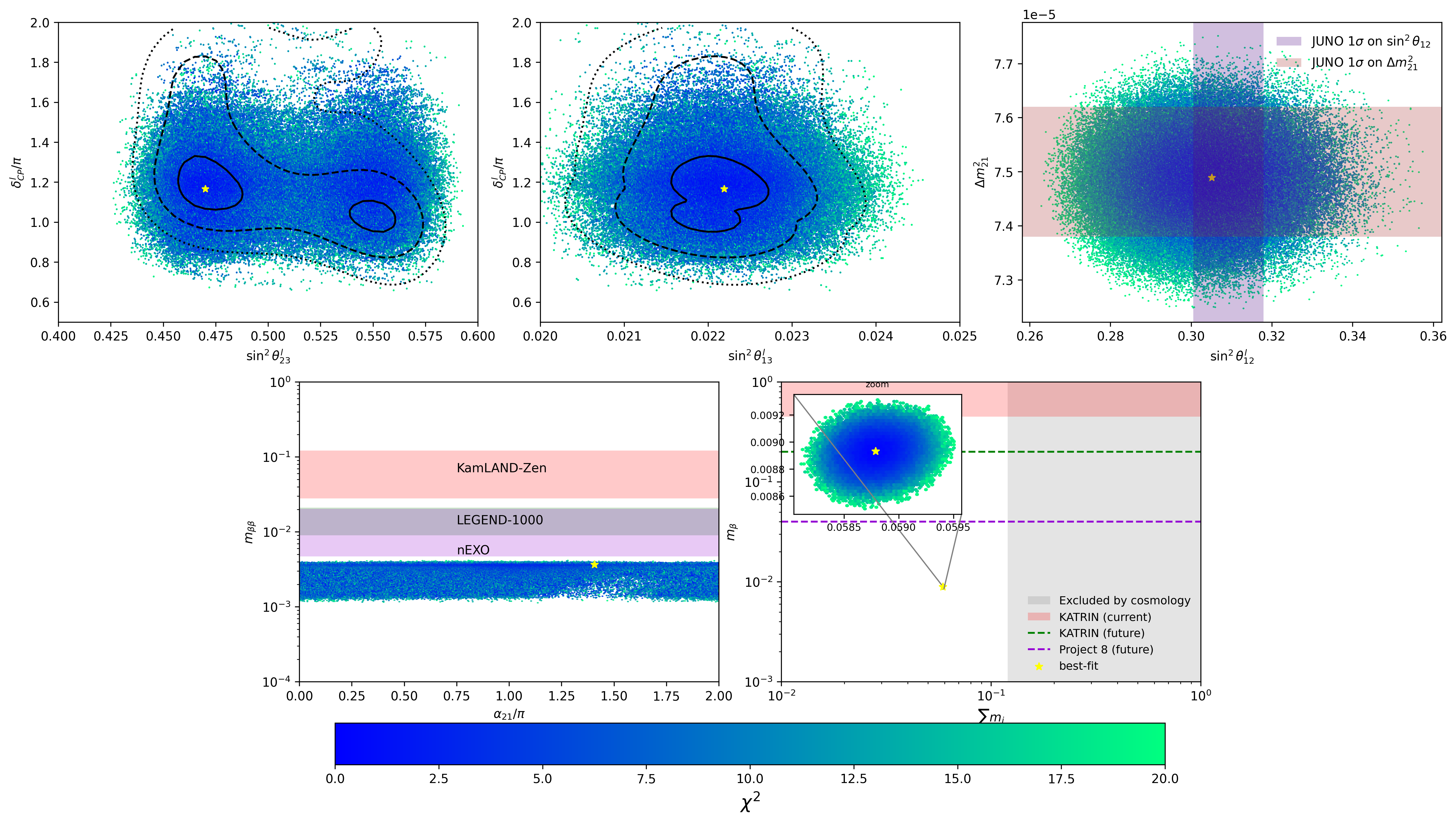}
\caption{Predicted correlations between {\it CP}-violating phases and mixing parameters of neutrinos for \textbf{BP 1} in the case of NO. The yellow star indicates the best-fitting point. Gray-shaded regions, excluded by cosmology, arise from the Planck constraint on the neutrino mass sum $\sum m_i$ \cite{Planck:2018vyg}. Shaded regions in the $m_\beta$ and $m_{\beta\beta}$ panels represent experimental limits from beta decay and $0\nu\beta\beta$ experiments.}
\label{fig5}
\end{figure}
In Table~\ref{results}, we present the predictions for the neutrino observables obtained using the benchmark points defined in Eqs. (\ref{bp1}) and (\ref{bp2}), which fix the scalar and VLL mass parameters that also enter the neutrino mass matrix. The remaining free parameters, namely the Yukawa couplings, are varied within the \texttt{FlavorPy} framework as part of the $\chi^2$ minimization procedure described above to reproduce the observed neutrino masses and mixing angles.
The table reports the best-fit values of the new Yukawa couplings $g_i$ and $\tilde{g_i}$ together with the model’s predictions for key observables, including fermion mass ratios, flavor-mixing parameters, the effective Majorana mass $m_{\beta\beta}$, the effective electron antineutrino mass $m_\beta$, and the light neutrino masses $m_{1,2,3}$. The minimum value $\chi^2$ for each benchmark is shown in the final row. The results are shown for both the normal and inverted neutrino mass orderings.
\begin{figure}[h]
\centering
\includegraphics[width=0.98 \columnwidth]{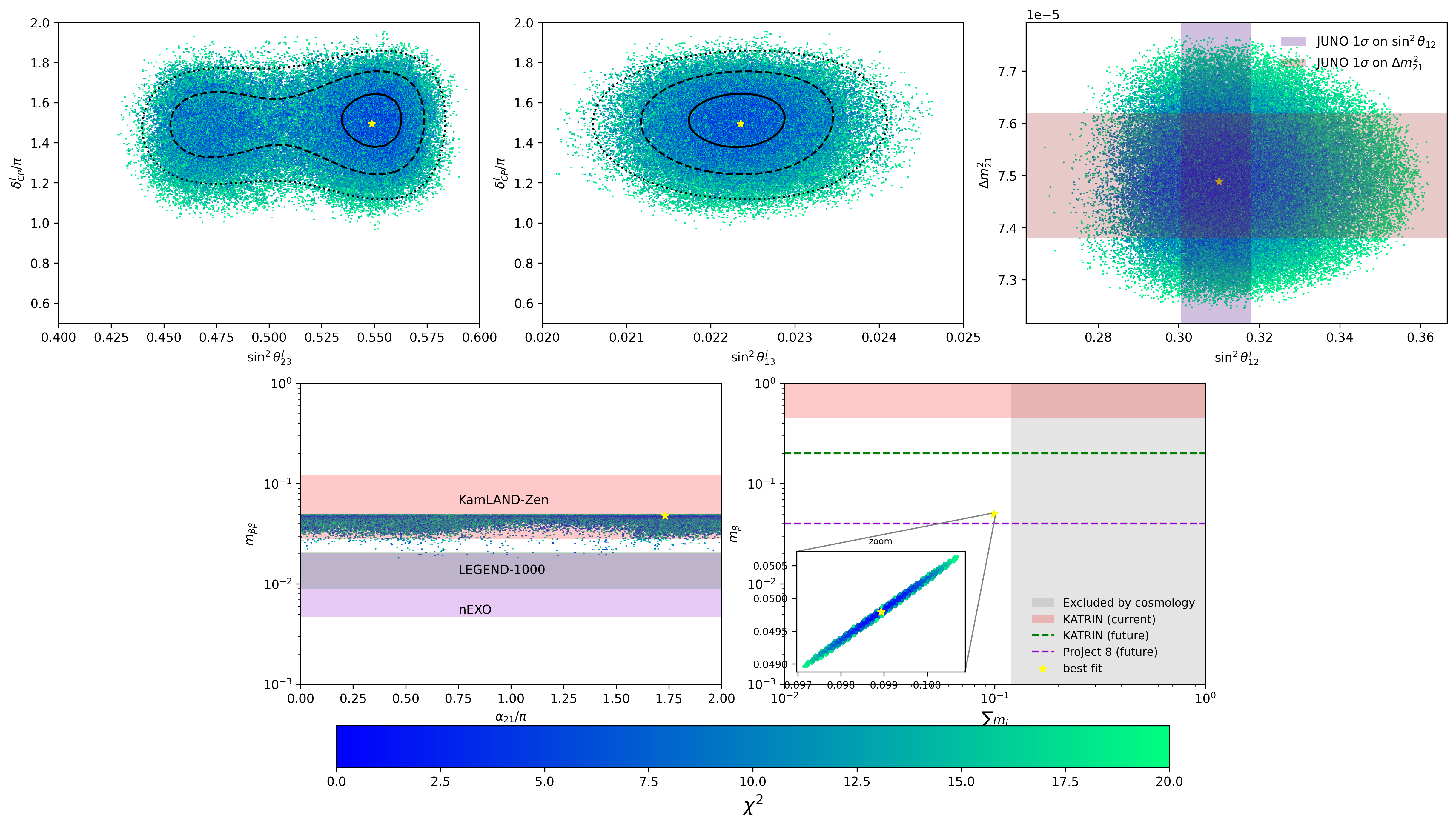}
\caption{Same as Fig. \ref{fig5} but for IO.}
\label{fig7}
\end{figure}
Both benchmarks are consistent with the experimental data within $1\sigma$. For BP 1, the total chi-square values are $\chi^2 \approx 0.011$ for NO and $\chi^2 \approx 0.057$ for IO, while for BP 2 they are $\chi^2 \approx 0.010$ for NO and $\chi^2 \approx 0.004$ for IO. Because BP 1 and BP 2 yield nearly indistinguishable best-fit values for the physical observables (see Table \ref{results}), showing both would result in visually redundant figures. We therefore restrict our presentation to BP 1 only, with the NO and IO cases displayed in Fig.~\ref{fig5} and Fig.~\ref{fig7}. In each figure, the first row contains three correlation panels, and the best-fit point is indicated by a yellow star. The contours denote confidence levels (CL) at $1\sigma$ (solid), $2\sigma$ (dashed), and $3\sigma$ (dash-dotted). The first panel shows the atmospheric mixing angle $\sin^2\theta_{23}^{l}$ versus the leptonic Dirac phase $\delta_{CP}^{l}$. The atmospheric angle lies in the lower octant ($\theta_{23} < 45^\circ$) for NO, and in the higher octant ($\theta_{23} > 45^\circ$) for IO, while both the mixing angle and $\delta_{CP}^{l}$ are accommodated within the $1\sigma$ region, where the best fit also resides. The second panel displays $\sin^2\theta_{13}^{l}$ against $\delta_{CP}^{l}$ where the preferred regions and the best-fit points fall within the $1\sigma$ contour, indicating excellent agreement for both the reactor angle and the Dirac phase. The third panel illustrates the correlation between the solar mass-squared difference, $\Delta m_{21}^2$, and the solar neutrino mixing angle, $\sin^2\theta_{12}$. The vertical and horizontal shaded bands denote the $1\sigma$ uncertainties of these parameters as reported by the JUNO experiment, which has very recently released its first measurements of $\Delta m_{21}^2$ and $\sin^2\theta_{12}$ \cite{JUNO:2025gmd}. Our model’s best-fit predictions for both parameters are in good agreement with the JUNO results.
\begin{figure}[h]
\centering
\includegraphics[width=0.93 \columnwidth]{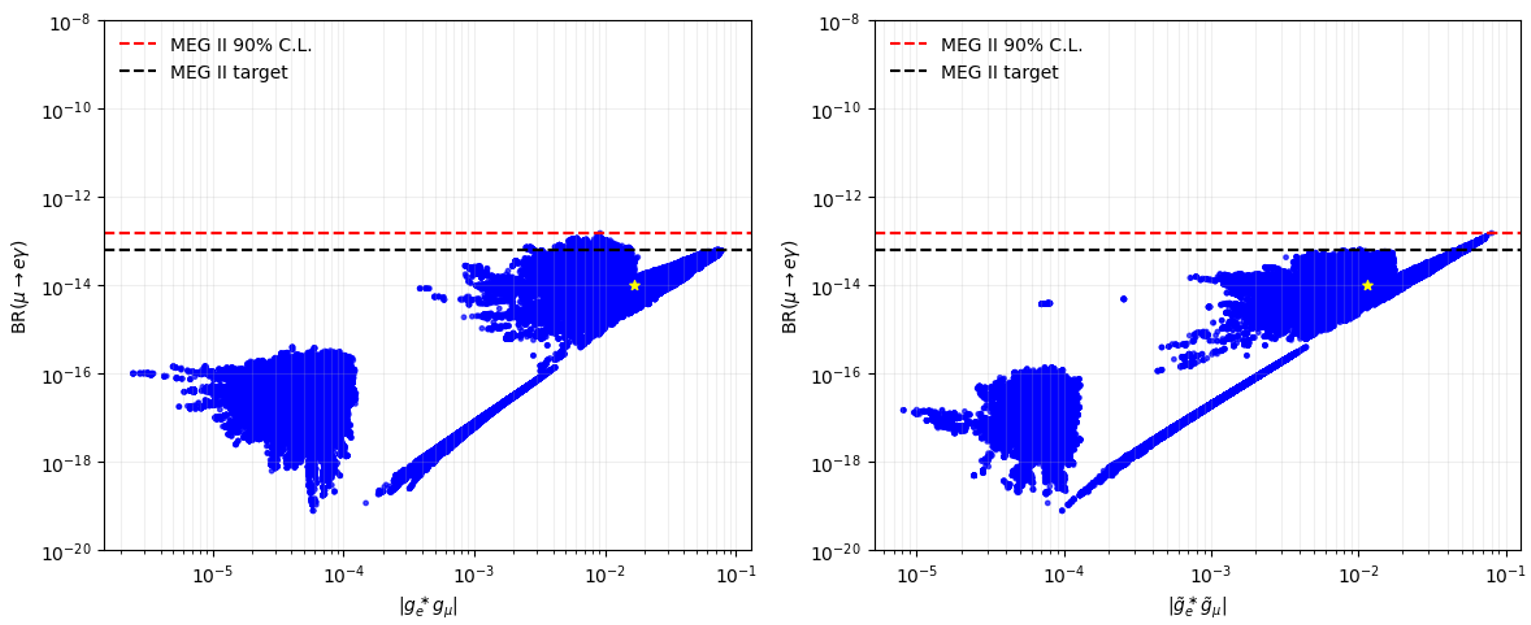}
\caption{Scatter plot of BR($\mu \rightarrow e \gamma$) versus $|g_e^\ast g_\mu|$ (left) and $|\tilde{g}_e^\ast \tilde{g}_\mu|$ for BP 1.}
\label{fig6}
\end{figure}
The second row of each figure presents two plots showing the model predictions for $m_{\beta\beta}$, $\alpha_{21}$, $\sum m_i$, and $m_\beta$. The corresponding best-fit values for both neutrino mass hierarchies and the two benchmark points are summarized in Table~\ref{results}. In particular, the left panel of Figs.~\ref{fig5} and~\ref{fig7} displays $m_{\beta\beta}$ as a function of the Majorana phase $\alpha_{21}$. For NO (Fig.~\ref{fig5}), the predicted best fit values of $m_{\beta\beta}$ lie below the current sensitivity of KamLAND-Zen~\cite{KamLAND-Zen:2024eml} and the projected reach of the upcoming $0\nu\beta\beta$ experiments LEGEND-1000~\cite{LEGEND:2021bnm} and nEXO~\cite{nEXO:2021ujk}, which aim to probe $m_{\beta\beta} \lesssim (4.7\text{--}20.3)~\mathrm{meV}$. The Majorana phase $\alpha_{21}$ remains unconstrained in this model, with its full range allowed. The predicted best fit values of the normalized phase, listed in Table~\ref{results}, correspond to a Majorana phase that allows for {\it CP} violation in the Majorana sector. For IO (Fig. \ref{fig7}), the best-fit value of the effective Majorana mass is $|m_{\beta\beta}| \simeq 47.9~\text{meV}$, which is compatible with the current KamLAND-Zen upper bound $m_{\beta\beta} < (28\text{--}122)~\text{meV}$, depending on nuclear matrix element uncertainties~\cite{KamLAND-Zen:2024eml}. In this scenario, the full range of $\alpha_{21}$ remains allowed. The right panels display the allowed regions in the ($m_\beta, \sum m_i$) plane. In our model, these parameters are highly constrained; therefore, we enlarge the area around the best-fit values to clearly illustrate the allowed regions as shown in both Fig. \ref{fig5} and Fig. \ref{fig7}. For NO, the predicted best-fit values are $m_\beta \simeq 8.9~\text{meV}$ and $\sum m_i \simeq 58.7~\text{meV}$. In this case, $m_\beta$ lies well below the projected sensitivity of the forthcoming Project~8 experiment ($\sim40~\text{meV}$), while $\sum m_i$ is comfortably below the current Planck upper bound, $\sum m_i < 120~\text{meV}$~\cite{Planck:2018vyg}. For IO, the predicted best-fit values are $m_\beta \simeq 49.7~\text{meV}$ and $\sum m_i \simeq 98.9~\text{meV}$. Here, $m_\beta$ exceeds the Project~8 sensitivity, potentially allowing the model prediction to be tested decisively, while $\sum m_i$ remains below the Planck limit but is closer to the upper bound than in the NO case.

\begin{figure}[h]
\centering
\includegraphics[width=0.93 \columnwidth]{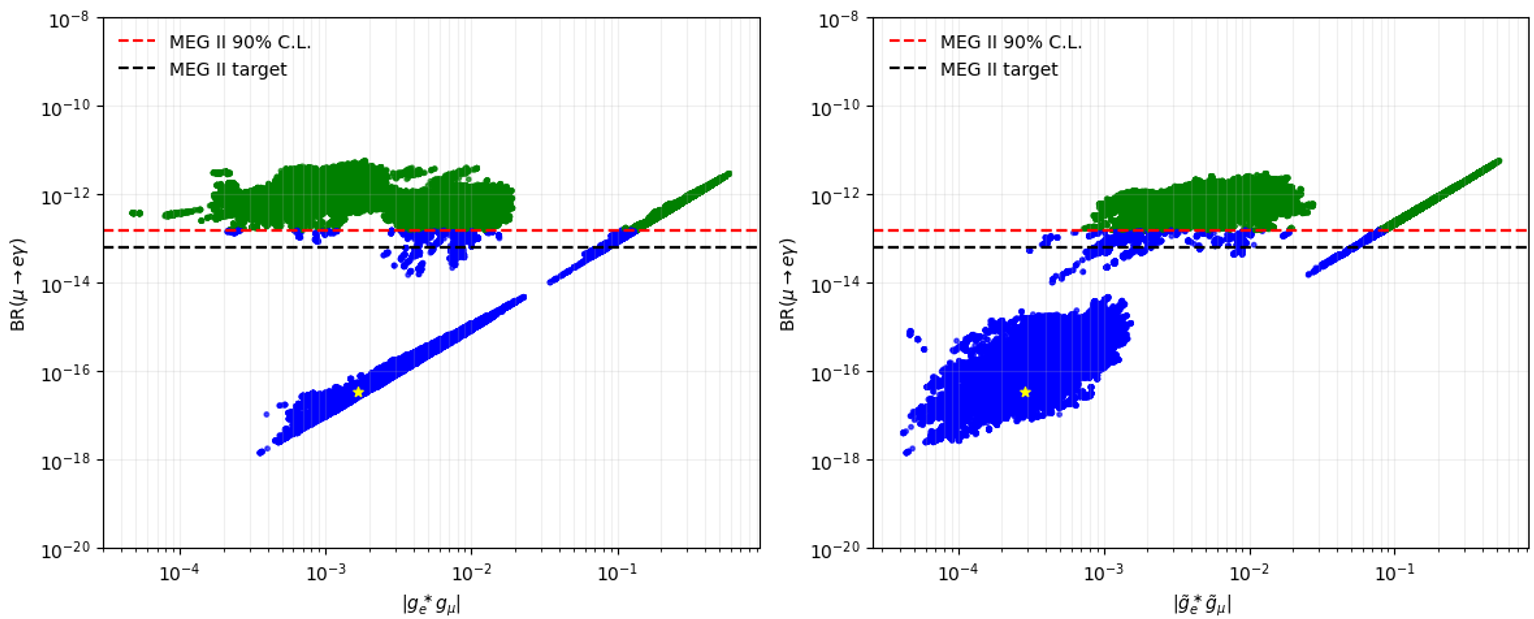}
\caption{Same as Fig. \ref{fig6} but for IO.}
\label{fig8}
\end{figure}

Regarding cLFV processes, we focus on the decay $\mu \rightarrow e \gamma$, whose branching ratio limit has been most stringently constrained by the MEG~II experiment \cite{MEGII:2025gzr}. The current bound given by $\text{BR}(\mu \rightarrow e \gamma) < 1.5 \times 10^{-13}$, represents one of the strongest probes of new physics in the charged-lepton sector. Moreover, MEG~II is expected to improve its sensitivity to approximately $6 \times 10^{-14}$ by the end of 2026 \cite{MEGII:2018kmf}. In Figs. \ref{fig6} and \ref{fig8}, we therefore plot $\text{BR}(\mu \rightarrow e \gamma)$ as a function of the product of the Yukawa couplings entering its expression; see Eqs. (\ref{BRemg}) and (\ref{eq:Aij}). This allows us to directly illustrate how the experimentally accessible branching ratio constrains the relevant combinations of model parameters. In both panels, parameter points consistent with the current MEG~II limits are shown in blue, while excluded points are shown in green. The yellow star denotes the best-fit point. For both mass hierarchies, the corresponding best-fit values lie below the current MEG~II sensitivity (red dashed lines) as well as below the projected sensitivity of the future MEG~II upgrade (black dashed lines).
%
%%%%%%%%%%%%%%%%%%%%%%%%%%%%%%%%%%%%%%%%%%%%%%%%%%%%%%%%%%%%%%%%
\section{Conclusions}
\label{sec:Conclusions} 
%%%%%%%%%%%%%%%%%%%%%%%%%%%%%%%%%%%%%%%%%%%%%%%%%%%%%%%%%%%%%%%%
%
In this work, we present a comprehensive framework for the generation of neutrino masses via a three-loop radiative mechanism intricately related to the presence of DM. The mechanism relies on four key ingredients. (i) Neutrino masses arise from asymmetric Yukawa couplings ($g_{ia} \neq \tilde g_{ia}$) involving new $SU(2)_L$ scalar doublets ($S_1$, $S_2$), vectorlike leptons with both left- and right-handed components ($F_{Li}$, $F_{Ri}$), and SM leptons. This asymmetry allows two nonzero neutrino masses to be generated even with a single generation of VLLs ($i=1$). As a result, the neutrino mass matrix takes the form $m^\nu_{ab} \propto (g_a \tilde{g}_b + \tilde{g}_a g_b)$, which ensures a nondegenerate neutrino mass matrix. (ii) The dependence on $\lambda_5 \propto m_H^2 - m_A^2$ ties neutrino masses to the mass splitting between the {\it CP}-even and {\it CP}-odd neutral scalars, requiring $m_H \neq m_A$. (iii) The dependence on $\lambda_7$, which mixes the singly charged fields is also crucial. Thus, $m^\nu_{ab} \propto \lambda_7 v^2 \propto s_\beta c_\beta(m_{H_1^+}^2-m_{H_2^+}^2)$, requiring $m_{H^+_1}\neq m_{H^+_2}$ for a nonzero neutrino mass. (iv) The Dirac mass insertion of the VLL, $m_E$, together with the three-loop integral $I_{3L}$, enters as $m^\nu_{ab} \propto m_E \times I_{3L}$. This term is crucial in modulating the dependence on $4m_E$. For smaller $m_E$, $I_{3L}$ is large and, conversely, for larger $m_E$, $I_{3L}$ is small, indicating a region where the product reaches a maximum, typically at the scale of the masses within the loop.

The mechanism described above provides a theoretical structure for the neutrino mass matrix, but its viability must be assessed against additional experimental constraints involving the same underlying parameters. To this end, we performed a numerical scan of the model’s parameter space, examining the neutrino mass spectrum, mixing angles, DM properties, and cLFV processes. The model accommodates current neutrino-oscillation data, satisfies stringent cLFV bounds—most notably on $\mu \to e\gamma$—and contains regions consistent with the Planck relic-density measurement $\Omega_{\rm DM} h^2$, present direct-detection limits from XENONnT and LZ, and the neutrino floor. This balance demonstrates the framework’s predictive capacity and its potential to yield experimentally testable signatures.

Our results highlight several promising directions for further investigation. Improved determinations of oscillation parameters, particularly the {\it CP}-violating phase $\delta_{CP}$, will critically test the predicted mixing pattern. Enhanced direct-detection sensitivity will further restrict the DM sector, while searches for cLFV processes—such as $\tau \to \mu\gamma$—can probe the Yukawa structure. Moreover, the collider phenomenology of the VLLs and additional scalars, as well as electroweak precision tests and further cLFV channels, remain to be explored in detail and will be addressed in future work within this framework.

Overall, the model offers a coherent and testable framework linking neutrino-mass generation to DM via asymmetric Yukawa interactions among new $SU(2)_L$ doublets, VLLs, and SM leptons. Forthcoming advances in neutrino-oscillation measurements, DM searches, and cLFV experiments will be essential for assessing and refining this scenario.
%%%%%%%%%%%%%%%%%%%%%%%%%%%%%%%%%%%%%%%%%%%%%%%%%%%%%%%%%%%%%%%%
\acknowledgments
%%%%%%%%%%%%%%%%%%%%%%%%%%%%%%%%%%%%%%%%%%%%%%%%%%%%%%%%%%%%%%%%
\noindent The work of M.A.L and S.N. is supported by the United Arab Emirates University (UAEU) under UPAR Grant No. 12S162. M.A.R. acknowledges support from Proyecto Interno USM PI LII 23 03 ``Neutrinos and Dark Matter: From Model Building to Phenomenology".
%%%%%%%%%%%%%%%%%%%%%%%%%%%%%%%%%%%%%%%%%%%%%%%%%%%%%%%%%%%%%%%%
\section*{Data availability}
%%%%%%%%%%%%%%%%%%%%%%%%%%%%%%%%%%%%%%%%%%%%%%%%%%%%%%%%%%%%%%%%
The data that support the findings of this article are not publicly available. The data are available from the authors upon reasonable request.

%%%%%%%%%%%%%%%%%%%%%%%%%%%%%%%%%%%%%%%%%%%%%%%%%%%%%%%%%%%%%%%%
\bibliographystyle{JHEP}
\bibliography{uv_knt}

@article{Cai:2017jrq,
    author = "Cai, Yi and Herrero-Garc{\'\i}a, Juan and Schmidt, Michael A. and Vicente, Avelino and Volkas, Raymond R.",
    title = "{From the trees to the forest: a review of radiative neutrino mass models}",
    eprint = "1706.08524",
    archivePrefix = "arXiv",
    primaryClass = "hep-ph",
    reportNumber = "ADP-17-29-T1035",
    doi = "10.3389/fphy.2017.00063",
    journal = "Front. in Phys.",
    volume = "5",
    pages = "63",
    year = "2017"
}

@article{Aoki:2008av,
    author = "Aoki, Mayumi and Kanemura, Shinya and Seto, Osamu",
    title = "{Neutrino mass, Dark Matter and Baryon Asymmetry via TeV-Scale Physics without Fine-Tuning}",
    eprint = "0807.0361",
    archivePrefix = "arXiv",
    primaryClass = "hep-ph",
    reportNumber = "UT-HET-011, IFT-UAM-CSIC-08-40",
    doi = "10.1103/PhysRevLett.102.051805",
    journal = "Phys. Rev. Lett.",
    volume = "102",
    pages = "051805",
    year = "2009"
}

@article{Kajiyama:2013lja,
    author = "Kajiyama, Yuji and Okada, Hiroshi and Yagyu, Kei",
    title = "{$T_7$ Flavor Model in Three Loop Seesaw and Higgs Phenomenology}",
    eprint = "1307.0480",
    archivePrefix = "arXiv",
    primaryClass = "hep-ph",
    reportNumber = "KIAS-P13034",
    doi = "10.1007/JHEP10(2013)196",
    journal = "JHEP",
    volume = "10",
    pages = "196",
    year = "2013"
}

@article{Ahriche:2014cda,
    author = "Ahriche, Amine and Chen, Chian-Shu and McDonald, Kristian L. and Nasri, Salah",
    title = "{Three-loop model of neutrino mass with dark matter}",
    eprint = "1404.2696",
    archivePrefix = "arXiv",
    primaryClass = "hep-ph",
    doi = "10.1103/PhysRevD.90.015024",
    journal = "Phys. Rev. D",
    volume = "90",
    pages = "015024",
    year = "2014"
}

@article{Ahriche:2014oda,
    author = "Ahriche, Amine and McDonald, Kristian L. and Nasri, Salah",
    title = "{A Model of Radiative Neutrino Mass: with or without Dark Matter}",
    eprint = "1404.5917",
    archivePrefix = "arXiv",
    primaryClass = "hep-ph",
    doi = "10.1007/JHEP10(2014)167",
    journal = "JHEP",
    volume = "10",
    pages = "167",
    year = "2014"
}

@article{Jin:2014glp,
    author = "Jin, Li-Gang and Tang, Rui and Zhang, Fei",
    title = "{A three-loop radiative neutrino mass model with dark matter}",
    eprint = "1501.02020",
    archivePrefix = "arXiv",
    primaryClass = "hep-ph",
    doi = "10.1016/j.physletb.2014.12.034",
    journal = "Phys. Lett. B",
    volume = "741",
    pages = "163--167",
    year = "2015"
}

@article{Hatanaka:2014tba,
    author = "Hatanaka, Hisaki and Nishiwaki, Kenji and Okada, Hiroshi and Orikasa, Yuta",
    title = "{A Three-Loop Neutrino Model with Global $U(1)$ Symmetry}",
    eprint = "1412.8664",
    archivePrefix = "arXiv",
    primaryClass = "hep-ph",
    reportNumber = "KIAS-P14079",
    doi = "10.1016/j.nuclphysb.2015.03.006",
    journal = "Nucl. Phys. B",
    volume = "894",
    pages = "268--283",
    year = "2015"
}

@article{Okada:2015hia,
    author = "Okada, Hiroshi and Yagyu, Kei",
    title = "{Three-loop neutrino mass model with doubly charged particles from isodoublets}",
    eprint = "1508.01046",
    archivePrefix = "arXiv",
    primaryClass = "hep-ph",
    doi = "10.1103/PhysRevD.93.013004",
    journal = "Phys. Rev. D",
    volume = "93",
    number = "1",
    pages = "013004",
    year = "2016"
}

@article{Nishiwaki:2015iqa,
    author = "Nishiwaki, Kenji and Okada, Hiroshi and Orikasa, Yuta",
    title = "{Three loop neutrino model with isolated $k^{\pm\pm}$}",
    eprint = "1507.02412",
    archivePrefix = "arXiv",
    primaryClass = "hep-ph",
    reportNumber = "KIAS-P15035",
    doi = "10.1103/PhysRevD.92.093013",
    journal = "Phys. Rev. D",
    volume = "92",
    number = "9",
    pages = "093013",
    year = "2015"
}

@article{Ahriche:2015wha,
    author = "Ahriche, Amine and McDonald, Kristian L. and Nasri, Salah and Toma, Takashi",
    title = "{A Model of Neutrino Mass and Dark Matter with an Accidental Symmetry}",
    eprint = "1504.05755",
    archivePrefix = "arXiv",
    primaryClass = "hep-ph",
    reportNumber = "LPT-ORSAY-15-29",
    doi = "10.1016/j.physletb.2015.05.031",
    journal = "Phys. Lett. B",
    volume = "746",
    pages = "430--435",
    year = "2015"
}

@article{Cheung:2017efc,
    author = "Cheung, Kingman and Nomura, Takaaki and Okada, Hiroshi",
    title = "{A Three-loop Neutrino Model with Leptoquark Triplet Scalars}",
    eprint = "1701.01080",
    archivePrefix = "arXiv",
    primaryClass = "hep-ph",
    reportNumber = "KIAS-P17002",
    doi = "10.1016/j.physletb.2017.03.021",
    journal = "Phys. Lett. B",
    volume = "768",
    pages = "359--364",
    year = "2017"
}

@article{Cepedello:2020lul,
    author = "Cepedello, Ricardo and Hirsch, Martin and Rocha-Mor{\'a}n, Paulina and Vicente, Avelino",
    title = "{Minimal 3-loop neutrino mass models and charged lepton flavor violation}",
    eprint = "2005.00015",
    archivePrefix = "arXiv",
    primaryClass = "hep-ph",
    reportNumber = "IFIC/20-17",
    doi = "10.1007/JHEP08(2020)067",
    journal = "JHEP",
    volume = "08",
    pages = "067",
    year = "2020"
}

@article{Super-Kamiokande:1998kpq,
    author = "Fukuda, Y. and others",
    collaboration = "Super-Kamiokande",
    title = "{Evidence for oscillation of atmospheric neutrinos}",
    eprint = "hep-ex/9807003",
    archivePrefix = "arXiv",
    reportNumber = "BU-98-17, ICRR-REPORT-422-98-18, UCI-98-8, KEK-PREPRINT-98-95, LSU-HEPA-5-98, UMD-98-003, SBHEP-98-5, TKU-PAP-98-06, TIT-HPE-98-09",
    doi = "10.1103/PhysRevLett.81.1562",
    journal = "Phys. Rev. Lett.",
    volume = "81",
    pages = "1562--1567",
    year = "1998"
}

@article{SNO:2002tuh,
    author = "Ahmad, Q. R. and others",
    collaboration = "SNO",
    title = "{Direct evidence for neutrino flavor transformation from neutral current interactions in the Sudbury Neutrino Observatory}",
    eprint = "nucl-ex/0204008",
    archivePrefix = "arXiv",
    doi = "10.1103/PhysRevLett.89.011301",
    journal = "Phys. Rev. Lett.",
    volume = "89",
    pages = "011301",
    year = "2002"
}

@article{Bertone:2004pz,
    author = "Bertone, Gianfranco and Hooper, Dan and Silk, Joseph",
    title = "{Particle dark matter: Evidence, candidates and constraints}",
    eprint = "hep-ph/0404175",
    archivePrefix = "arXiv",
    reportNumber = "FERMILAB-PUB-04-047-A",
    doi = "10.1016/j.physrep.2004.08.031",
    journal = "Phys. Rept.",
    volume = "405",
    pages = "279--390",
    year = "2005"
}

@article{Mohapatra:2005wg,
    author = "Mohapatra, R. N. and others",
    title = "{Theory of Neutrinos: A White Paper}",
    eprint = "hep-ph/0510213",
    archivePrefix = "arXiv",
    reportNumber = "FERMILAB-TM-2342-T, SLAC-PUB-11622",
    doi = "10.1088/0034-4885/70/11/R02",
    journal = "Rept. Prog. Phys.",
    volume = "70",
    pages = "1757--1867",
    year = "2007"
}

@article{Chen:2006vn,
      author         = "Chen, Chian-Shu and Geng, C. Q. and Ng, J. N.",
      title          = "{Unconventional Neutrino Mass Generation, Neutrinoless
                        Double Beta Decays, and Collider Phenomenology}",
      journal        = "Phys. Rev.",
      volume         = "D75",
      year           = "2007",
      pages          = "053004",
      doi            = "10.1103/PhysRevD.75.053004",
      eprint         = "hep-ph/0610118",
      archivePrefix  = "arXiv",
      primaryClass   = "hep-ph",
      SLACcitation   = "%%CITATION = HEP-PH/0610118;%%"
}

@article{delAguila:2011gr,
      author         = "del Aguila, Francisco and Aparici, Alberto and
                        Bhattacharya, Subhaditya and Santamaria, Arcadi and Wudka,
                        Jose",
      title          = "{A realistic model of neutrino masses with a large
                        neutrinoless double beta decay rate}",
      journal        = "JHEP",
      volume         = "05",
      year           = "2012",
      pages          = "133",
      doi            = "10.1007/JHEP05(2012)133",
      eprint         = "1111.6960",
      archivePrefix  = "arXiv",
      primaryClass   = "hep-ph",
      reportNumber   = "CAFPE-165-11, UG-FT-295-11, FTUV-11-1128, IFIC-11-65,
                        UCRHEP-T512",
      SLACcitation   = "%%CITATION = ARXIV:1111.6960;%%"
}

@article{Gustafsson:2012vj,
      author         = "Gustafsson, Michael and No, Jose Miguel and Rivera,
                        Maximiliano A.",
      title          = "{Predictive Model for Radiatively Induced Neutrino Masses
                        and Mixings with Dark Matter}",
      journal        = "Phys. Rev. Lett.",
      volume         = "110",
      year           = "2013",
      number         = "21",
      pages          = "211802",
      doi            = "10.1103/PhysRevLett.110.211802,
                        10.1103/PhysRevLett.112.259902",
      note           = "[Erratum: Phys. Rev. Lett.112,no.25,259902(2014)]",
      eprint         = "1212.4806",
      archivePrefix  = "arXiv",
      primaryClass   = "hep-ph",
      reportNumber   = "ULB-TH-12-23, USM-TH-310",
      SLACcitation   = "%%CITATION = ARXIV:1212.4806;%%"
}

@article{Alcaide:2017xoe,
      author         = "Alcaide, Julien and Das, Dipankar and Santamaria, Arcadi",
      title          = "{A model of neutrino mass and dark matter with large
                        neutrinoless double beta decay}",
      journal        = "JHEP",
      volume         = "04",
      year           = "2017",
      pages          = "049",
      doi            = "10.1007/JHEP04(2017)049",
      eprint         = "1701.01402",
      archivePrefix  = "arXiv",
      primaryClass   = "hep-ph",
      reportNumber   = "FTUV-17-0104, IFIC-16-97",
      SLACcitation   = "%%CITATION = ARXIV:1701.01402;%%"
}

@article{Lavoura:2003xp,
    author = "Lavoura, L.",
    title = "{General formulae for f(1) ---{\ensuremath{>}} f(2) gamma}",
    eprint = "hep-ph/0302221",
    archivePrefix = "arXiv",
    doi = "10.1140/epjc/s2003-01212-7",
    journal = "Eur. Phys. J. C",
    volume = "29",
    pages = "191--195",
    year = "2003"
}

@article{ParticleDataGroup:2024cfk,
    author = "Navas, S. and others",
    collaboration = "Particle Data Group",
    title = "{Review of particle physics}",
    doi = "10.1103/PhysRevD.110.030001",
    journal = "Phys. Rev. D",
    volume = "110",
    number = "3",
    pages = "030001",
    year = "2024"
}

@article{Esteban:2024eli,
    author = "Esteban, Ivan and Gonzalez-Garcia, M. C. and Maltoni, Michele and Martinez-Soler, Ivan and Pinheiro, Jo\~ao Paulo and Schwetz, Thomas",
    title = "{NuFit-6.0: updated global analysis of three-flavor neutrino oscillations}",
    eprint = "2410.05380",
    archivePrefix = "arXiv",
    primaryClass = "hep-ph",
    reportNumber = "IFT-UAM/CSIC-24-140, YITP-SB-2024-24, IPPP/24/64, IPPP/24/64, IFT-UAM/CSIC-24-140, YITP-SB-2024-24",
    doi = "10.1007/JHEP12(2024)216",
    journal = "JHEP",
    volume = "12",
    pages = "216",
    year = "2024"
}

@article{Xing:2007fb,
    author = "Xing, Zhi-zhong and Zhang, He and Zhou, Shun",
    title = "{Updated Values of Running Quark and Lepton Masses}",
    eprint = "0712.1419",
    archivePrefix = "arXiv",
    primaryClass = "hep-ph",
    doi = "10.1103/PhysRevD.77.113016",
    journal = "Phys. Rev. D",
    volume = "77",
    pages = "113016",
    year = "2008"
}

@software{FlavorPy,
  author        = {Baur, Alexander},
  title         = "{FlavorPy}",
  year          = {2024},
  publisher     = {Zenodo},
  version       = {v0.2.0},
  doi           = {10.5281/zenodo.11060597},
  url           = "\url{https://doi.org/10.5281/zenodo.11060597}"
}

@article{Chen:2014ska,
    author = "Chen, Chian-Shu and McDonald, Kristian L. and Nasri, Salah",
    title = "{A Class of Three-Loop Models with Neutrino Mass and Dark Matter}",
    eprint = "1404.6033",
    archivePrefix = "arXiv",
    primaryClass = "hep-ph",
    doi = "10.1016/j.physletb.2014.05.082",
    journal = "Phys. Lett. B",
    volume = "734",
    pages = "388--393",
    year = "2014"
}

@article{Krauss:2002px,
      author         = "Krauss, Lawrence M. and Nasri, Salah and Trodden, Mark",
      title          = "{A Model for neutrino masses and dark matter}",
      journal        = "Phys. Rev.",
      volume         = "D67",
      year           = "2003",
      pages          = "085002",
      doi            = "10.1103/PhysRevD.67.085002",
      eprint         = "hep-ph/0210389",
      archivePrefix  = "arXiv",
      primaryClass   = "hep-ph",
      reportNumber   = "SU-DP-02-10-3, SU-4252-770, CWRU-P41-02",
      SLACcitation   = "%%CITATION = HEP-PH/0210389;%%"
}

@article{Borowka:2017idc,
    author = "Borowka, S. and Heinrich, G. and Jahn, S. and Jones, S.P. and Kerner, M. and Schlenk, J. and Zirke, T.",
    title = "{pySecDec: a toolbox for the numerical evaluation of multi-scale integrals}",
    eprint = "1703.09692",
    archivePrefix = "arXiv",
    primaryClass = "hep-ph",
    reportNumber = "MPP-2017-42, CERN-TH-2017-063, IPPP-17-24",
    doi = "10.1016/j.cpc.2017.09.015",
    journal = "Comput. Phys. Commun.",
    volume = "222",
    pages = "313--326",
    year = "2018"
}

@article{Gustafsson:2014vpa,
    author = "Gustafsson, Michael and No, Jose M. and Rivera, Maximiliano A.",
    title = "{Radiative neutrino mass generation linked to neutrino mixing and 0$\nu$$\beta$$\beta$-decay predictions}",
    eprint = "1402.0515",
    archivePrefix = "arXiv",
    primaryClass = "hep-ph",
    reportNumber = "ULB-TH-14-04, USM-TH-320",
    doi = "10.1103/PhysRevD.90.013012",
    journal = "Phys. Rev. D",
    volume = "90",
    number = "1",
    pages = "013012",
    year = "2014"
}

@article{Kannike:2012pe,
    author = "Kannike, Kristjan",
    title = "{Vacuum Stability Conditions From Copositivity Criteria}",
    eprint = "1205.3781",
    archivePrefix = "arXiv",
    primaryClass = "hep-ph",
    doi = "10.1140/epjc/s10052-012-2093-z",
    journal = "Eur. Phys. J. C",
    volume = "72",
    pages = "2093",
    year = "2012"
}

@article{JUNO:2025gmd,
    author = "Abusleme, Angel and others",
    collaboration = "JUNO",
    title = "{First measurement of reactor neutrino oscillations at JUNO}",
    eprint = "2511.14593",
    archivePrefix = "arXiv",
    primaryClass = "hep-ex",
    month = "11",
    year = "2025"
}

@article{Ma:2006km,
    author = "Ma, Ernest",
    title = "{Verifiable radiative seesaw mechanism of neutrino mass and dark matter}",
    eprint = "hep-ph/0601225",
    archivePrefix = "arXiv",
    reportNumber = "UCRHEP-T403",
    doi = "10.1103/PhysRevD.73.077301",
    journal = "Phys. Rev. D",
    volume = "73",
    pages = "077301",
    year = "2006"
}

@article{Babu:2001ex,
      author         = "Babu, K. S. and Leung, Chung Ngoc",
      title          = "{Classification of effective neutrino mass operators}",
      journal        = "Nucl. Phys.",
      volume         = "B619",
      year           = "2001",
      pages          = "667-689",
      doi            = "10.1016/S0550-3213(01)00504-1",
      eprint         = "hep-ph/0106054",
      archivePrefix  = "arXiv",
      primaryClass   = "hep-ph",
      reportNumber   = "OSU-HEP-01-02, UDHEP-02-01",
      SLACcitation   = "%%CITATION = HEP-PH/0106054;%%"
}

@article{deGouvea:2007qla,
      author         = "de Gouvea, Andre and Jenkins, James",
      title          = "{A Survey of Lepton Number Violation Via Effective
                        Operators}",
      journal        = "Phys. Rev.",
      volume         = "D77",
      year           = "2008",
      pages          = "013008",
      doi            = "10.1103/PhysRevD.77.013008",
      eprint         = "0708.1344",
      archivePrefix  = "arXiv",
      primaryClass   = "hep-ph",
      reportNumber   = "NUHEP-TH-07-10",
      SLACcitation   = "%%CITATION = ARXIV:0708.1344;%%"
}

@article{Gustafsson:2020bou,
    author = "Gustafsson, Michael and No, José Miguel and Rivera, Maximiliano A.",
    title = "{Lepton number violating operators with standard model gauge fields: A survey of neutrino masses from 3-loops and their link to dark matter}",
    eprint = "2006.13564",
    archivePrefix = "arXiv",
    primaryClass = "hep-ph",
    doi = "10.1007/JHEP11(2020)070",
    journal = "JHEP",
    volume = "11",
    pages = "070",
    year = "2020"
}

@article{Cirelli:2005uq,
    author = "Cirelli, Marco and Fornengo, Nicolao and Strumia, Alessandro",
    title = "{Minimal dark matter}",
    eprint = "hep-ph/0512090",
    archivePrefix = "arXiv",
    reportNumber = "DFTT40-2005, IFUP-TH-2005-34",
    doi = "10.1016/j.nuclphysb.2006.07.012",
    journal = "Nucl. Phys. B",
    volume = "753",
    pages = "178--194",
    year = "2006"
}

@article{Bhattacharya:2018fus,
    author = "Bhattacharya, Subhaditya and Ghosh, Purusottam and Sahoo, Nirakar and Sahu, Narendra",
    title = "{Mini Review on Vector-Like Leptonic Dark Matter, Neutrino Mass, and Collider Signatures}",
    eprint = "1812.06505",
    archivePrefix = "arXiv",
    primaryClass = "hep-ph",
    doi = "10.3389/fphy.2019.00080",
    journal = "Front. in Phys.",
    volume = "7",
    pages = "80",
    year = "2019"
}

@article{XENON:2023cxc,
    author = "Aprile, E. and others",
    collaboration = "XENON",
    title = "{First Dark Matter Search with Nuclear Recoils from the XENONnT Experiment}",
    eprint = "2303.14729",
    archivePrefix = "arXiv",
    primaryClass = "hep-ex",
    doi = "10.1103/PhysRevLett.131.041003",
    journal = "Phys. Rev. Lett.",
    volume = "131",
    number = "4",
    pages = "041003",
    year = "2023"
}

@article{LZ:2024zvo,
    author = "Aalbers, J. and others",
    collaboration = "LZ",
    title = "{Dark Matter Search Results from 4.2{\,}{\,}Tonne-Years of Exposure of the LUX-ZEPLIN (LZ) Experiment}",
    eprint = "2410.17036",
    archivePrefix = "arXiv",
    primaryClass = "hep-ex",
    reportNumber = "FERMILAB-PUB-24-0796-V",
    doi = "10.1103/4dyc-z8zf",
    journal = "Phys. Rev. Lett.",
    volume = "135",
    number = "1",
    pages = "011802",
    year = "2025"
}

@article{Planck:2018vyg,
    author = "Aghanim, N. and others",
    collaboration = "Planck",
    title = "{Planck 2018 results. VI. Cosmological parameters}",
    eprint = "1807.06209",
    archivePrefix = "arXiv",
    primaryClass = "astro-ph.CO",
    doi = "10.1051/0004-6361/201833910",
    journal = "Astron. Astrophys.",
    volume = "641",
    pages = "A6",
    year = "2020",
    note = "[Erratum: Astron.Astrophys. 652, C4 (2021)]"
}

@article{OHare:2021utq,
    author = "O'Hare, Ciaran A. J.",
    title = "{New Definition of the Neutrino Floor for Direct Dark Matter Searches}",
    eprint = "2109.03116",
    archivePrefix = "arXiv",
    primaryClass = "hep-ph",
    doi = "10.1103/PhysRevLett.127.251802",
    journal = "Phys. Rev. Lett.",
    volume = "127",
    number = "25",
    pages = "251802",
    year = "2021"
}

@article{Alloul:2013bka,
    author = "Alloul, Adam and Christensen, Neil D. and Degrande, C{\'e}line and Duhr, Claude and Fuks, Benjamin",
    title = "{FeynRules  2.0 - A complete toolbox for tree-level phenomenology}",
    eprint = "1310.1921",
    archivePrefix = "arXiv",
    primaryClass = "hep-ph",
    reportNumber = "CERN-PH-TH-2013-239, MCNET-13-14, IPPP-13-71, DCPT-13-142, PITT-PACC-1308",
    doi = "10.1016/j.cpc.2014.04.012",
    journal = "Comput. Phys. Commun.",
    volume = "185",
    pages = "2250--2300",
    year = "2014"
}

@article{Belyaev:2012qa,
    author = "Belyaev, Alexander and Christensen, Neil D. and Pukhov, Alexander",
    title = "{CalcHEP 3.4 for collider physics within and beyond the Standard Model}",
    eprint = "1207.6082",
    archivePrefix = "arXiv",
    primaryClass = "hep-ph",
    reportNumber = "PITT-PACC-1209",
    doi = "10.1016/j.cpc.2013.01.014",
    journal = "Comput. Phys. Commun.",
    volume = "184",
    pages = "1729--1769",
    year = "2013"
}

@article{Belanger:2013oya,
    author = "Belanger, G. and Boudjema, F. and Pukhov, A. and Semenov, A.",
    title = "{micrOMEGAs$\_$3: A program for calculating dark matter observables}",
    eprint = "1305.0237",
    archivePrefix = "arXiv",
    primaryClass = "hep-ph",
    reportNumber = "LAPTH-023-13",
    doi = "10.1016/j.cpc.2013.10.016",
    journal = "Comput. Phys. Commun.",
    volume = "185",
    pages = "960--985",
    year = "2014"
}

@article{Alguero:2023zol,
    author = "Alguero, G. and Belanger, G. and Boudjema, F. and Chakraborti, S. and Goudelis, A. and Kraml, S. and Mjallal, A. and Pukhov, A.",
    title = "{micrOMEGAs 6.0: N-component dark matter}",
    eprint = "2312.14894",
    archivePrefix = "arXiv",
    primaryClass = "hep-ph",
    doi = "10.1016/j.cpc.2024.109133",
    journal = "Comput. Phys. Commun.",
    volume = "299",
    pages = "109133",
    year = "2024"
}

@article{Kanemura:2010sh,
    author = "Kanemura, Shinya and Matsumoto, Shigeki and Nabeshima, Takehiro and Okada, Nobuchika",
    title = "{Can WIMP Dark Matter overcome the Nightmare Scenario?}",
    eprint = "1005.5651",
    archivePrefix = "arXiv",
    primaryClass = "hep-ph",
    reportNumber = "UT-HET-039",
    doi = "10.1103/PhysRevD.82.055026",
    journal = "Phys. Rev. D",
    volume = "82",
    pages = "055026",
    year = "2010"
}

@article{Alarcon:2011zs,
    author = "Alarcon, J. M. and Martin Camalich, J. and Oller, J. A.",
    title = "{The chiral representation of the $\pi N$ scattering amplitude and the pion-nucleon sigma term}",
    eprint = "1110.3797",
    archivePrefix = "arXiv",
    primaryClass = "hep-ph",
    doi = "10.1103/PhysRevD.85.051503",
    journal = "Phys. Rev. D",
    volume = "85",
    pages = "051503",
    year = "2012"
}

@article{Alarcon:2012nr,
    author = "Alarcon, J. M. and Geng, L. S. and Martin Camalich, J. and Oller, J. A.",
    title = "{The strangeness content of the nucleon from effective field theory and phenomenology}",
    eprint = "1209.2870",
    archivePrefix = "arXiv",
    primaryClass = "hep-ph",
    doi = "10.1016/j.physletb.2014.01.065",
    journal = "Phys. Lett. B",
    volume = "730",
    pages = "342--346",
    year = "2014"
}

@article{Mambrini:2011ik,
    author = "Mambrini, Y.",
    title = "{Higgs searches and singlet scalar dark matter: Combined constraints from XENON 100 and the LHC}",
    eprint = "1108.0671",
    archivePrefix = "arXiv",
    primaryClass = "hep-ph",
    doi = "10.1103/PhysRevD.84.115017",
    journal = "Phys. Rev. D",
    volume = "84",
    pages = "115017",
    year = "2011"
}

@article{ATLAS:2012yve,
    author = "Aad, Georges and others",
    collaboration = "ATLAS",
    title = "{Observation of a new particle in the search for the Standard Model Higgs boson with the ATLAS detector at the LHC}",
    eprint = "1207.7214",
    archivePrefix = "arXiv",
    primaryClass = "hep-ex",
    reportNumber = "CERN-PH-EP-2012-218",
    doi = "10.1016/j.physletb.2012.08.020",
    journal = "Phys. Lett. B",
    volume = "716",
    pages = "1--29",
    year = "2012"
}

@article{CMS:2013btf,
    author = "Chatrchyan, Serguei and others",
    collaboration = "CMS",
    title = "{Observation of a New Boson with Mass Near 125 GeV in $pp$ Collisions at $\sqrt{s}$ = 7 and 8 TeV}",
    eprint = "1303.4571",
    archivePrefix = "arXiv",
    primaryClass = "hep-ex",
    reportNumber = "CMS-HIG-12-036, CERN-PH-EP-2013-035",
    doi = "10.1007/JHEP06(2013)081",
    journal = "JHEP",
    volume = "06",
    pages = "081",
    year = "2013"
}

@article{Deshpande:1977rw,
    author = "Deshpande, Nilendra G. and Ma, Ernest",
    title = "{Pattern of Symmetry Breaking with Two Higgs Doublets}",
    reportNumber = "OITS-81",
    doi = "10.1103/PhysRevD.18.2574",
    journal = "Phys. Rev. D",
    volume = "18",
    pages = "2574",
    year = "1978"
}

@article{Barbieri:2006dq,
    author = "Barbieri, Riccardo and Hall, Lawrence J. and Rychkov, Vyacheslav S.",
    title = "{Improved naturalness with a heavy Higgs: An Alternative road to LHC physics}",
    eprint = "hep-ph/0603188",
    archivePrefix = "arXiv",
    reportNumber = "UCB-PTH-06-04, LBNL-59894",
    doi = "10.1103/PhysRevD.74.015007",
    journal = "Phys. Rev. D",
    volume = "74",
    pages = "015007",
    year = "2006"
}

@software{newville2016lmfit,
  author={Newville, Matthew and Stensitzki, Till and Allen, Daniel B and Rawlik, Michal and Ingargiola, Antonino and Nelson, Andrew},
  title="{LMFIT: Non-linear least-square minimization and curve-fitting for Python}",
  journal={Astrophysics Source Code Library},
  pages={ascl--1606},
  year={2016},
  url = "\url{https://doi.org/10.5281/zenodo.16175987}"
}

@article{KamLAND-Zen:2024eml,
    author = "Abe, S. and others",
    collaboration = "KamLAND-Zen",
    title = "{Search for Majorana Neutrinos with the Complete KamLAND-Zen Dataset}",
    eprint = "2406.11438",
    archivePrefix = "arXiv",
    primaryClass = "hep-ex",
    month = "6",
    year = "2024"
}

@article{LEGEND:2021bnm,
    author = "Abgrall, N. and others",
    collaboration = "LEGEND",
    title = "{The Large Enriched Germanium Experiment for Neutrinoless $\beta\beta$ Decay}: {LEGEND-1000 Preconceptual Design Report}",
    eprint = "2107.11462",
    archivePrefix = "arXiv",
    primaryClass = "physics.ins-det",
    month = "7",
    year = "2021"
}

@article{nEXO:2021ujk,
    author = "Adhikari, G. and others",
    collaboration = "nEXO",
    title = "{nEXO: neutrinoless double beta decay search beyond 10$^{28}$ year half-life sensitivity}",
    eprint = "2106.16243",
    archivePrefix = "arXiv",
    primaryClass = "nucl-ex",
    doi = "10.1088/1361-6471/ac3631",
    journal = "J. Phys. G",
    volume = "49",
    number = "1",
    pages = "015104",
    year = "2022"
}

@article{MEGII:2025gzr,
    author = "Afanaciev, K. and others",
    collaboration = "MEG II",
    title = "{New limit on the $\mu^{+} \to e^{+} \gamma$ decay with the MEG II experiment}",
    eprint = "2504.15711",
    archivePrefix = "arXiv",
    primaryClass = "hep-ex",
    doi = "10.1140/epjc/s10052-025-14906-3",
    journal = "Eur. Phys. J. C",
    volume = "85",
    number = "10",
    pages = "1177",
    year = "2025"
}

@article{MEGII:2018kmf,
    author = "Baldini, A. M. and others",
    collaboration = "MEG II",
    title = "{The design of the MEG II experiment}",
    eprint = "1801.04688",
    archivePrefix = "arXiv",
    primaryClass = "physics.ins-det",
    doi = "10.1140/epjc/s10052-018-5845-6",
    journal = "Eur. Phys. J. C",
    volume = "78",
    number = "5",
    pages = "380",
    year = "2018"
}
%%%%%%%%%%%%%%%%%%%%%%%%%%%%%%%%%%%%%%%%%%%%%%%%%%%%%%%%%%%%%%%%%%%%%%%%%%%%
\end{document}